\documentclass[a4paper,11pt]{article}
\pdfoutput=1 % if your are submitting a pdflatex (i.e. if you have
\usepackage{jcappub} % for details on the use of the package, please
\usepackage{amssymb}
\usepackage{graphicx}
\usepackage[T1]{fontenc} % if needed

\newcommand{\be}{\begin{equation}}
\newcommand{\ee}{\end{equation}}

\newcommand{\lp}{\left(}
\newcommand{\rp}{\right)}
\newcommand{\lb}{\left[}
\newcommand{\rb}{\right]}

\newcommand{\ab}{\bar{a}}
\newcommand{\cb}{\bar{c}}
\begin{document}
\title{\boldmath Generalised Quadratic Curvature, Non-Local Infrared Modifications of Gravity and Newtonian Potentials}
%\title{\boldmath non-local infrared modifications of gravity and the post-Newtonian limit}

%% %simple case: 2 authors, same institution
%% \author{A. Uthor}
%% \author{and A. Nother Author}
%% \affiliation{Institution,\\Address, Country}

% more complex case: 4 authors, 3 institutions, 2 footnotes
\author[a]{Aindri\'u Conroy,}
\author[b]{Tomi Koivisto,}
%\author[c,3]{Alexey S. Koshelev\note{Also at Some University.}}
\author[a]{Anupam Mazumdar}
\author[a]{and Ali Teimouri}

% The "\note" macro will give a warning: "Ignoring empty anchor..."
% you can safely ignore it.

\affiliation[a]{Consortium for Fundamental Physics, Lancaster University, Lancaster, LA1 4YB, UK}
\affiliation[b]{Nordita, KTH Royal Institute of Technology and Stockholm University, Roslagstullsbacken 23,
SE-10691 Stockholm, Sweden}

% e-mail addresses: one for each author, in the same order as the authors
%\emailAdd{tbiswas@loyno.edu}
\emailAdd{a.conroy@lancaster.ac.uk}
\emailAdd{tomik@astro.uio.no}
%\emailAdd{alexey.koshelev@vub.ac.be}
\emailAdd{a.mazumdar@lancaster.ac.uk}
\emailAdd{a.teimouri@lancaster.ac.uk}

%\preprint{NORDITA-2014-73}

\abstract{Metric theories of gravity are studied, beginning with a general action that is quadratic in curvature and allows arbitrary {\it inverse powers} of the 
d'Alembertian 
operator, resulting in infrared non-local extensions of general relativity. The field equations are derived in full generality and their consistency is checked by verifying the Bianchi identities.  The weak-field limit is computed and a straightforward algorithm is presented to infer the post-Newtonian corrections directly from the action.
This is then applied to various infrared gravity models including non-local $Rf(R/ \Box)$ dark energy  and non-local massive gravity models. 
Generically, the Newtonian potentials are not identical and deviate from the $1/r$ behaviour at large distances. However, the former does not occur in a specific class of theories that does not introduce additional degrees of freedom in flat spacetime. A new non-local model within this class is proposed, defined by the exponential of the inverse d'Alembertian. This model exhibits novel features, such as the weakening of the gravity in the infrared, suggesting de-gravitation of the cosmological constant.  }

\maketitle
\flushbottom
\renewcommand\[{\begin{equation}}
\renewcommand\]{\end{equation}}
\newcommand{\al}{\alpha}
\newcommand{\bt}{\beta}
\newcommand{\ga}{\gamma}
\newcommand{\da}{\delta}
\newcommand{\n}{\nabla}
\newcommand{\ba}{\begin{eqnarray}}
\newcommand{\ea}{\end{eqnarray}}
\newcommand{\LF}{\left(}
\newcommand{\RF}{\right)}
\newcommand{\LT}{\left[}
\newcommand{\RT}{\right]}
\newcommand{\Ld}{\left.}
\newcommand{\Rd}{\right.}
\newcommand{\cO}{{\cal O}}
\newcommand{\cF}{{\cal F}}
\newcommand{\Ra}{\Rightarrow}
\newcommand{\bl}{\begin{aligned}}
\newcommand{\el}{\end{aligned}}
\newcommand{\nn}{\nonumber}

\section{Introduction}

It has been known for some time that higher derivative theory of gravity can be renormalisable but only at the cost of unitarity \cite{Stelle:1976gc}. There is some evidence that in an infinite-order higher derivative, i.e. non-local theory of gravity, one can avoid the issue of ghosts and other pathologies while recovering general relativity (GR) at low energies. The propagator for the most general metric theory was recently derived in \cite{Biswas:2011ar}, see also e.g. \cite{Tomboulis:1997gg,Modesto:2011kw,Biswas:2013kla}. In non-locally improved theories, gravity becomes weak in the ultraviolet (UV), yielding non-singular black hole, gravitational wave and cosmological solutions. Many aspects of ghost-free and singularity-free gravity have been studied in the context of early Universe cosmology \cite{Biswas:2005qr,Biswas:2006bs,Modesto:2010uh,Biswas:2010zk,Biswas:2011ar,Biswas:2012bp,Biswas:2013dry,Chialva:2014rla}.

However, non-local operators in the gravity sector may also play a role in the infrared (IR). In particular, they could filter out the contribution of the cosmological constant to the gravitating energy density, possibly providing the key to solving
one of the most notorious problems in physics \cite{ArkaniHamed:2002fu}, see also \cite{Dvali:2007kt,Barvinsky:2003kg,Nojiri:2010pw, Zhang:2011uv,Barvinsky:2012ts}. Cosmological implications of non-local terms in the gravity action, such as $R\Box^{-1} R$, have been investigated for instance motivated by the possibility to render the Euclidean action finite \cite{Wetterich:1997bz}. More recently, many studies of cosmologies in infrared, non-locally modified gravity models have been undertaken, stimulated by the problems of dark matter and dark energy \cite{Koivisto:2009jn,Woodard:2014iga}. In this paper we will focus on the infrared non-localities of gravity in a very generic manner. We consider metric covariant quadratic curvature theories of gravity, allowing arbitrary (including up to infinite order) inverse derivatives in the action.
Typically the corrections are then suppressed by some IR scale, which we shall denote as $M$ everywhere in the following.

Deser and Woodard proposed non-local corrections to the gravity action in the form $Rf(R/\Box)$ \cite{Deser:2007jk}.
By such means one could perhaps address some of the fine-tuning problems of dark energy: the curvature scalar $R$
is negligible with respect to the radiation density early on, which might help to understand why the corrections become significant only during the
matter dominated epoch, and on the other hand, $R\Box$ being a dimensionless combination,
modifications at the scale of dark energy might be generated without introduction of tiny mass
scales into the theory \cite{Koivisto:2008xfa}. The $Rf(R/\Box)$ models have been studied extensively~\cite{Deffayet:2009ca,Bronnikov:2009az,Nojiri:2010pw, Zhang:2011uv,Elizalde:2011su,Elizalde:2012ja,Elizalde:2013dlt,Deser:2013uya}, and in particular, cosmological perturbations have been analysed in Refs.~\cite{Koivisto:2008dh,Nesseris:2009jf,Park:2012cp,Dodelson:2013sma}, with the conclusion that the current structure formation data clearly favours GR over the $Rf(R/\Box)$ models when the background evolution of the latter is fixed to be identical to the $\Lambda$CDM. In fact, this model is ruled out at the confidence level of several sigmas \cite{Dodelson:2013sma}. Furthermore, in more general models, involving tensorial non-local terms such as $R_{\mu\nu}(M^2/ \Box^2) R^{\mu\nu}$, an additional, potentially dangerous, growing mode appears in the cosmological perturbations \cite{Foffa:2013vma,Ferreira:2013tqn}.

However, this does not render all non-local gravity dark energy models incompatible with cosmological constraints, as demonstrated by two interesting viable examples recently put forward by Maggiore {\it et al}. One model was defined by a $g_{\mu\nu}R/\Box$ term added to the Einstein field equations \cite{Maggiore:2013mea,Foffa:2013sma,Kehagias:2014sda} and its viability was verified against a number of large scale structure data~\cite{Nesseris:2014mea}. The other model was defined by adding a $R \Box^{-2} R$ modification to the Lagrangian \cite{Maggiore:2014sia}. This introduces a single-parameter alternative to the $\Lambda$CDM cosmology, whose perturbation evolution has also been  shown to produce a matter power spectrum that matches well with current measurements \cite{Dirian:2014ara}. The possibility that dark matter could be a manifestation of a non-local deviation from Einstein's gravity has been investigated by several authors as well \cite{Soussa:2003vv,Hehl:2008eu,Barvinsky:2011hd,Barvinsky:2011rk,Deffayet:2011sk,Arraut:2013qra,Deffayet:2014lba,Woodard:2014wia}. 
%The non-linear structure formation and the constraints on these models of gravity arising from the high redshift Universe remain to be studied~\cite{2013MNRAS.434.1486D,2014arXiv1405.4862D}.

Non-local gravity poses many theoretical and technical issues. The initial value problem has been considered \cite{Barnaby:2008tc,Barnaby:2010kx} and practical methods of solving non-local systems of differential equations, such as the diffusion equation approach, have been developed \cite{Aref'eva:2007uk,Mulryne:2008iq,Nunes:2008ta,Calcagni:2010ab}.
In the non-local framework, the graviton can be given a mass without introducing an additional metric \cite{Jaccard:2013gla,Modesto:2013jea}. One could thus speculate that non-local gravity may describe massive ghost-free gravity, once the additional metric has been integrated out along the lines of \cite{Hassan:2013pca}.
Indeed it has been argued that the infrared non-local gravity models proposed to date can only be taken as phenomenological effective theories \cite{Foffa:2013sma,Maggiore:2013mea,Tsamis:2014hra}.
Two techniques have been employed to generate causal and conserved field equations: either by varying an invariant non-local effective action and then enforcing causality by the ad hoc replacement of any advanced Green's function with its retarded counterpart, or by introducing causal non-locality into a general ansatz for the field equations and then enforcing conservation. These approaches are implemented in the two examples of dark energy models mentioned above, respectively. 

In this paper we adopt the first approach: our starting point is a general non-local action. The most general linear equations were analysed in \cite{Biswas:2011ar} and \cite{Biswas:2013cha}, where the authors have presented the most general non-linear field equations for non-local gravity up to quadratic order in the curvature, with the aim to understand the UV properties of gravity. Here, we proceed by extending the analysis into the IR. We begin, in Section \ref{sec:action} by describing the quadratic action and deriving the field equations in full generality. Due to the complicated nature of these calculations, it is useful to perform a consistency check by verifying that they satisfy the Bianchi identities. This non-trivial calculation is outlined briefly. In Section \ref{sec:limit} we consider the weak-field limit of these theories and present an algorithm to compute the Newtonian potentials. These can be very useful in determining the observational viability of a theory at the level of astrophysics and classical tests of gravity within the Solar system. In the following section, we apply our formalism to specific models, by way of three examples, before proposing a new model featuring the exponential of the inverse d'Alembertian operator. We give some concluding remarks in Section \ref{sec:conclusions} and some technical details have been confined to the appendices, along with a scalar presentation of a restricted class of models (Appendix \ref{sec:scalar}).

%\cite{Sandstad:2013oja}.

\section{General Quadratic Action and Field Equations} \label{sec:action}

In a pioneering paper, Schmidt considered the field equations in quadratic-curvature gravity theories of arbitrarily high derivative order \cite{Schmidt:1990dh}, then restricting to modifications of GR in terms of the Ricci scalar. Only recently, the full nonlinear analysis was generalised to arbitrary curvature terms  \cite{Biswas:2013cha}, motivated by the progress made with such theories in Ref.~\cite{Biswas:2011ar} where it was shown that gravity in the UV can be made asymptotically-free without violating basic principles of physics such as unitarity and general covariance. Here we extend the action into the IR, and describe it as follows
\[
\label{action}
S=\int d^{4}x\frac{\sqrt{-g}}{2}\left(M_{P}^{2}R+R\bar{{\cal F}_{1}}(\Box)R+R^{\mu\nu}\bar{{\cal F}_{2}}(\Box)R_{\mu\nu}+ C^{\mu\nu\lambda\sigma}\bar{{\cal F}_{3}}(\Box)C_{\mu\nu\lambda\sigma}\right)\,,
  \]
 where
 \[
 \label{F}
 \bar{{\cal F}}_{i}(\Box)=\sum_{n=1}^{\infty}f_{i_{-n}}\Box^{-n}\qquad\mbox{with }i=(1,2,3)\,.
 \]
 with $f_{i_{-n}}={\tilde f}_{i_{-n}}M^{2n}$, where ${\tilde f}_{i_{-n}}$ is a constant to ensure correct dimensionality and $M$ is some infrared mass scale. To derive the field equations from this action, we need to first understand the properties of the inverse d'Alembertian operator under variations of the metric.
% Repeated integration by parts gives us the identity 
% \[
% T\delta(\Box^{-n})S=\sum^{n+1}_{m=0} \Box^{-m}T\delta(\Box)\Box^{m-n-1}S
% \]
% where $T$ and $S$ are tensors of any type.

 \subsection{Variation of the Inverse D'Alembertian}
 \label{2-1}
We compute the equations of motion of \eqref{action} by straightforwardly taking the variation of the action. We note that most of the terms, can be found by adhering to the prescription given in \cite{Biswas:2013cha}. However, one particular brand of term requires more attention, namely the $\delta\bar{{\cal F}}_{i}(\Box)$-type terms, which we shall discuss briefly below.

Following the prescription of \cite{Foffa:2013vma}, \cite{Dirian:2014xoa}, in order to preserve causality and the conservation of the energy-momentum tensor, we consider only solutions with vanishing homogenous solution, namely where
\[
\label{delta-1}
\delta(\Box^{-1})S=-\Box^{-1}\delta(\Box)\Box^{-1}S
\]
Further details of this can be found in Appendix \ref{sec:app2-1}. Applying the product rule, we find
\[
\label{delta-n}
\delta(\Box^{-n})S=\sum_{m=0}^{n-1}\Box^{-m}\delta(\Box^{-1})\Box^{-n+m+1}S
 \]
Using the defintion of the function \eqref{F} and substituting \eqref{delta-1} into the above equation, we find the general form of $\delta\bar{{\cal F}}_{i}(\Box)S$ to be
\[
\delta\bar{{\cal F}}_{i}(\Box)S=-\sum_{n=1}^{\infty}f_{i_{-n}}\sum_{m=0}^{n-1}\Box^{-m-1}\delta(\Box)\Box^{-n+m}S
 \]
%\\\\
%Let $S$ and $T$ be tensors of any type. As should be clear, we note that the variation of the identity vanishes. We may then write
%\[
%\delta(\Box^{n}\Box^{-n})S=0=\delta(\Box^{n})\Box^{-n}S+\Box^{n}\delta(\Box^{-n})S
%\] 
%
%implying
%\[
%\Box^{n}\delta(\Box^{-n})S=-\delta(\Box^{n})\Box^{-n}S
% \]
%
%so that
%\[
%T\delta(\Box^{-n})S=-T\Box^{-n}\delta(\Box^{n})\Box^{-n}S
% \]
%
%Recall from \cite{Biswas:2013cha} that
%\[
%T\delta(\Box^{n})S=\sum_{m=0}^{n-1}T\Box^{m}\delta(\Box)\Box^{n-m-1}S
% \]
%
%which can be verified by simply expanding out powers of $\Box$
% . Upon substituting, we find
%\[
%T\delta(\Box^{-n})S=-\sum_{m=0}^{n-1}T\Box^{m-n}\delta(\Box)\Box^{-m-1}S
% \]
%
%and from the definition of $\bar{{\cal F}}_{i}(\Box)$
% , we may write
%\[
%T\delta{\cal \bar{F}}_{i}(\Box)S=-\sum_{n=1}^{\infty}f_{i_{n}}\sum_{m=0}^{n-1}T\Box^{m-n}\delta(\Box)\Box^{-m-1}S
% \]
%
%Using integration by parts, we can express this as follows
%\[
%T\delta{\cal \bar{F}}_{i}(\Box)S=-\sum_{n=1}^{\infty}f_{i_{n}}\sum_{m=0}^{n-1}\Box^{m-n}T\delta(\Box)\Box^{-m-1}S
% \]
%
%or equivalently
%\[
%T\delta{\cal \bar{F}}_{i}(\Box)S=-\sum_{n=1}^{\infty}f_{i_{n}}\sum_{m=0}^{n-1}T^{(m-n)}\delta(\Box)S^{(-m-1)}
% \]

Finally, from \cite{Biswas:2013cha} , we know how $\delta(\Box)$
  acts upon the curvature scalar, Ricci tensor and Weyl tensor and following the prescription therein, we can read off the equations of motion. Details of these variational terms are given in Appendix \ref{sec:var}.
 
\subsection{Equations of Motion}

The field equations are:
 \[
 \label{eomir}
\begin{aligned}T_{\alpha\beta} & =M_{P}^{2}G_{\alpha\beta}+2G_{\alpha\beta}\bar{{\cal F}_{1}}(\Box)R+\frac{1}{2}g_{\alpha\beta}R\bar{{\cal F}_{1}}(\Box)R-2\left(\nabla_{\alpha}\nabla_{\beta}-g_{\alpha\beta}\square\right)\bar{{\cal F}_{1}}(\Box)R\\
 & +\Theta_{\alpha\beta}^{1}-\frac{1}{2}g_{\alpha\beta}\left(\Theta_{\sigma}^{1\sigma}+\bar{\Theta}^{1}\right)+2R_{\alpha\sigma}\bar{{\cal F}_{2}}(\Box)R_{\;\beta}^{\sigma}\\
 & -\frac{1}{2}g_{\alpha\beta}R_{\nu}^{\mu}\bar{{\cal F}_{2}}(\Box)R_{\mu}^{\nu}-2\nabla_{\sigma}\nabla_{\beta}\bar{{\cal F}_{2}}(\Box)R_{\alpha}^{\;\sigma}+\square\bar{{\cal F}_{2}}(\Box)R_{\alpha\beta}+g_{\alpha\beta}\nabla_{\mu}\nabla_{\nu}\bar{{\cal F}_{2}}(\Box)R^{\mu\nu}\\
 & +\Theta_{\alpha\beta}^{2}-\frac{1}{2}g_{\alpha\beta}\left(\Theta_{\sigma}^{2\sigma}+\bar{\Theta}^{2}\right)+2{\cal E}_{\alpha\beta}^{2}\\
 & -\frac{1}{2}g_{\alpha\beta}C^{\mu\nu\lambda\sigma}\bar{{\cal F}_{3}}(\Box)C_{\mu\nu\lambda\sigma}+2C_{\alpha\mu\nu\sigma}\bar{{\cal {\cal F}}_{3}}(\square)C_{\beta}^{\;\mu\nu\sigma}-2\left(R_{\mu\nu}+2\nabla_{\mu}\nabla_{\nu}\right)\bar{{\cal {\cal F}}_{3}}(\square)C_{\beta\alpha}^{\;\;\mu\nu}\\
 & +\Theta_{\alpha\beta}^{3}-\frac{1}{2}g_{\alpha\beta}\left(\Theta_{\sigma}^{3\sigma}+\bar{\Theta}^{3}\right)+4{\cal E}_{\alpha\beta}^{3}\,, 
\end{aligned}
     \]
where we have defined the following tensors
\[
\nn
\Theta_{\alpha\beta}^{1}=\sum_{n=1}^{\infty}f_{1_{-n}}\sum_{l=0}^{n-1}\nabla_{\beta}R^{(-l-1)}\nabla_{\alpha}R^{(-n+l)},\qquad\bar{\Theta}^{1}=\sum_{n=1}^{\infty}f_{1_{-n}}\sum_{l=0}^{n-1}R^{(-l-1)}R^{(-n+l+1)},
 \] 
\[
\nn
\Theta_{\alpha\beta}^{2}=\sum_{n=1}^{\infty}f_{2_{-n}}\sum_{l=0}^{n-1}\nabla_{\alpha}R_{\nu}^{\mu(-l-1)}\nabla_{\beta}R_{\mu}^{\nu(-n+l)},\quad\bar{\Theta}^{2}=\sum_{n=1}^{\infty}f_{2_{-n}}\sum_{l=0}^{n-1}R_{\nu}^{\mu(-l-1)}R_{\mu}^{\nu(-n+l+1)}\,,
  \]
\[
\nn
{\cal E}_{\alpha\beta}^{2}=\frac{1}{2}\sum_{n=1}^{\infty}f_{2_{-n}}\sum_{l=0}^{n-1}\nabla_{\nu}\left(R_{\;\sigma}^{\nu(-l-1)}\nabla_{(\alpha}R_{\beta)}^{\;\sigma(-n+l)}-\nabla_{\alpha}R_{\;\sigma}^{\nu(-l-1)}R_{\beta)}^{\;\sigma(-n+l)}\right)\,,
  \]
\[
\nn
\Theta_{\alpha\beta}^{3}=\sum_{n=1}^{\infty}f_{3_{-n}}\sum_{l=0}^{n-1}\nabla_{\alpha}C_{\;\nu\lambda\sigma}^{\mu(-l-1)}\nabla_{\beta}C_{\mu}^{\;\nu\lambda\sigma(-n+l)},\quad\bar{\Theta}^{3}=\sum_{n=1}^{\infty}f_{3_{-n}}\sum_{l=0}^{n-1}C_{\;\nu\lambda\sigma}^{\mu(-l-1)}C_{\mu}^{\;\nu\lambda\sigma(-n+l+1)}\,,
   \]
\[
\label{tensors}
{\cal E}_{\alpha\beta}^{3}=\frac{1}{2}\sum_{n=1}^{\infty}f_{3_{-n}}\sum_{l=0}^{n-1}\nabla_{\nu}\left(C_{\;\;\;\sigma\mu}^{\lambda\nu(-l-1)}\nabla_{(\alpha}C_{|\lambda|\beta)}^{\;\;\;\sigma\mu(-n+l)}-\nabla_{(\alpha}C_{\;\;\;\sigma\mu}^{\lambda\nu(-l-1)}C_{\lambda\beta)}^{\;\;\sigma\mu(-n+l)}\right)\,.
  \]

 \subsection{Bianchi Identity Test}

 The stress-energy tensor of any minimally coupled diffeomorphism invariant gravitational action must be conserved,
 \[
 \nabla^\beta T_{\alpha\beta}=0\,.
 \]
 Furthermore, it should be noted that the Bianchi identities should hold for each 'part' of the action \eqref{action}, with the first 'part' comprised of the Einstein-Hilbert action and the following three accounting for the $\bar{{\cal F}}_1(\Box)$, $\bar{{\cal F}}_2(\Box)$ and $\bar{{\cal F}}_3(\Box)$ sections, as each of these sections are independent of each other. Clearly, the Einstein-Hilbert action satisfies the Bianchi identity as the Einstein tensor satisfies $\nabla^\beta G_{\alpha\beta}=0$.

 Let us begin with the piece
 \[
 \label{S1}
 S_1=\int d^4 x \sqrt{-g}\biggl(R{\bar{\cal F}}_1(\Box) R\biggr)\,.
 \]
% which can be applied to the entire action, with an amendment to \eqref{comm} depending on the tensor-type in question. 
Expanding the tensors given in \eqref{tensors}, we may write the equation of motion for \eqref{S1} as follows:
 \[
\begin{aligned}T^1_{\alpha\beta} & =2G_{\alpha\beta}\bar{{\cal F}_{1}}(\Box)R+\frac{1}{2}g_{\alpha\beta}R\bar{{\cal F}_{1}}(\Box)R-2\left(\nabla_{\alpha}\nabla_{\beta}-g_{\alpha\beta}\square\right)\bar{{\cal F}_{1}}(\Box)R\\
 & +\Theta_{\alpha\beta}^{1}-\frac{1}{2}g_{\alpha\beta}\left(\Theta_{\sigma}^{1\sigma}+\bar{\Theta}^{1}\right)\,.
\end{aligned}
  \]
We then take the covariant derivative and cancel like terms
\[
\begin{aligned}T_{\alpha\beta}^{1;\beta} & =\frac{1}{2}\nabla_{\alpha}R\bar{{\cal F}_{1}}(\Box)R+2R_{\alpha\sigma}\nabla^{\sigma}\bar{{\cal F}_{1}}(\Box)R-\frac{1}{2}R\nabla_{\alpha}\bar{{\cal F}_{1}}(\Box)R-2\nabla^{\sigma}\nabla_{\alpha}\nabla_{\sigma}\bar{{\cal F}_{1}}(\Box)R+2\nabla_{\alpha}\square\bar{{\cal F}_{1}}(\Box)R\\
 & +\sum_{n=1}^{\infty}f_{1_{-n}}\sum_{l=0}^{n-1}\biggl[\Box R^{(-l-1)}\nabla_{\alpha}R^{(l-n)}+\frac{1}{2}\nabla_{\sigma}R^{(-l-1)}\nabla^{\sigma}\nabla_{\alpha}R^{(l-n)}-\frac{1}{2}\nabla_{\alpha}\nabla_{\sigma}R^{(-l-1)}\nabla^{\sigma}R^{(l-n)}\\
 & -\frac{1}{2}\nabla_{\alpha}R^{(-l-1)}R^{(-n+l+1)}-\frac{1}{2}R^{(-l-1)}\nabla_{\alpha}R^{(-n+l+1)}\biggr]\,.
\end{aligned}
     \]
Next we use
\[
\label{comm}
[\nabla_{a},\nabla_{b}]\lambda^{c}=R_{\; dab}^{c}\lambda^{d}
  \]
to find
\[
\nabla_{\sigma}\nabla_{\alpha}\nabla^{\sigma}\bar{{\cal F}_{1}}(\Box)R=\nabla_{\alpha}\Box\bar{{\cal F}_{1}}(\Box)R+R_{\sigma\alpha}\nabla^{\sigma}\bar{{\cal F}_{1}}(\Box)R
  \]
and substitute to obtain
\[
\begin{aligned}T_{\alpha\beta}^{1;\beta} & =\frac{1}{2}\nabla_{\alpha}R\bar{{\cal F}_{1}}(\Box)R-\frac{1}{2}R\nabla_{\alpha}\bar{{\cal F}_{1}}(\Box)R\\
 & +\sum_{n=1}^{\infty}f_{1_{-n}}\sum_{l=0}^{n-1}\biggl[R^{(-l)}\nabla_{\alpha}R^{(l-n)}+\frac{1}{2}\nabla_{\sigma}R^{(-l-1)}\nabla^{\sigma}\nabla_{\alpha}R^{(l-n)}-\frac{1}{2}\nabla_{\alpha}\nabla_{\sigma}R^{(-l-1)}\nabla^{\sigma}R^{(l-n)}\\
 & -\frac{1}{2}\nabla_{\alpha}R^{(-l-1)}R^{(-n+l+1)}-\frac{1}{2}R^{(-l-1)}\nabla_{\alpha}R^{(-n+l+1)}\biggr]\,.
\end{aligned}
      \]
All remaining terms will cancel by noting that
\[
\int d^{4}x\sqrt{-g}\sum_{n=1}^{\infty}\sum_{m=0}^{n-1}A^{(m)}B^{(n)}=\int d^{4}x\sqrt{-g}\sum_{n=1}^{\infty}\sum_{m=0}^{n-1}A^{(n)}B^{(m)}
  \]
and thus the Bianchi identities are satisfied. A similar method may be used to test for the Bianchi identities of the entire action using the general formula
\[
\begin{aligned}
{[}\nabla_{\rho},\nabla_{\sigma}{]}X_{\;\;\;\;\;\nu_{1}...\nu_{l}}^{\mu_{1}...\mu_{k}}	&=	R_{\;\lambda\rho\sigma}^{\mu_{1}}X_{\;\;\;\;\;\nu_{1}...\nu_{l}}^{\lambda\mu_{2}...\mu_{k}}+R_{\;\lambda\rho\sigma}^{\mu_{2}}X_{\;\;\;\;\;\;\nu_{1}...\nu_{l}}^{\mu_{1}\lambda\mu_{3}...\mu_{k}}+...
		\\&-R_{\;\nu_{1}\rho\sigma}^{\lambda}X_{\;\;\;\;\;\lambda...\nu_{l}}^{\mu_{1}...\mu_{k}}-R_{\;\nu_{2}\rho\sigma}^{\lambda}X_{\;\;\;\;\;\nu_{1}\lambda\nu_{3}...\nu_{l}}^{\mu_{1}...\mu_{k}}-...
 \end{aligned}
 \]
 Further details are given in Appendix \ref{sec:Bianchi} for the $\bar{{\cal F}}_2(\Box)$ and $\bar{{\cal F}}_3(\Box)$ pieces of the action.

\section{Weak-Field Limit} \label{sec:limit}

In order to make a step towards understanding the physical implications of the theories analysed in Section \ref{sec:examples} and to make contact with observations, let us consider the weak-field limit of the general field equations.

From $g_{\mu\nu}=\eta_{\mu\nu}+h_{\mu\nu}$
 and the definition of the Christoffel symbols and the Riemann tensor, one can find the weak-field limit of the Riemann tensor, Ricci tensor and curvature scalar,
\[ 
\nonumber
 R_{\rho\mu\sigma\nu}=\frac{1}{2}\left(\partial_{\sigma}\partial_{\mu}h_{\rho\nu}+\partial_{\nu}\partial_{\rho}h_{\mu\sigma}-\partial_{\nu}\partial_{\mu}h_{\rho\sigma}-\partial_{\sigma}\partial_{\rho}h_{\mu\nu}\right)\,,
 \]
\[
\nonumber
R_{\mu\nu}=\frac{1}{2}\left(\partial^{\sigma}\partial_{\mu}h_{\sigma\nu}+\partial_{\nu}\partial_{\sigma}h_{\mu}^{\;\sigma}-\partial_{\nu}\partial_{\mu}h-\Box h_{\mu\nu}\right)\,,
 \]
\[
\label{Rs}
R=\partial_{\mu}\partial_{\nu}h^{\mu\nu}-\square h\,.
 \]
as well as the Weyl tensor which is somewhat lengthier and is given in appendix \ref{sec:weyl}.

In the weak-field limit, we may discount terms of order $h^{2}$ and higher. With this in mind, the equation of motion \eqref{eomir} reduces significantly,
\[
\begin{aligned}T_{\alpha\beta} & =M_{P}^{2}G_{\alpha\beta}-2\left(\nabla_{\alpha}\nabla_{\beta}-\eta_{\alpha\beta}\square\right)\bar{{\cal F}_{1}}(\Box)R-2\nabla_{\sigma}\nabla_{\beta}\bar{{\cal F}_{2}}(\Box)R_{\alpha}^{\;\sigma}\\
 & +\square\bar{{\cal F}_{2}}(\Box)R_{\alpha\beta}+\eta_{\alpha\beta}\nabla_{\mu}\nabla_{\nu}\bar{{\cal F}_{2}}(\Box)R^{\mu\nu}-4\nabla_{\mu}\nabla_{\nu}\bar{{\cal {\cal F}}_{3}}(\square)C_{\beta\alpha}^{\;\;\mu\nu}\,,
\end{aligned}
  \]
into which we can then substitute the above values for the Riemann tensor, Ricci tensor and curvature scalar \eqref{Rs} to obtain %in order to describe the weak-field corresponding to the IR given below
\[
\label{lineom}
\begin{aligned}T_{\alpha\beta} & =-\left[1+\frac{1}{2}\bar{{\cal F}_{2}}(\Box)\square+\bar{{\cal F}_{3}}(\Box)\Box\right]\square h_{\alpha\beta}\\
 & -\left[-1-\frac{1}{2}\bar{{\cal F}_{2}}(\Box)\Box-\bar{{\cal F}_{3}}(\Box)\Box\right]\partial_{\sigma}\left(\partial_{\alpha}h_{\;\beta}^{\sigma}+\partial_{\beta}h_{\alpha}^{\;\sigma}\right)\\
 & -\left[1-2\bar{{\cal F}_{1}}(\Box)\square-\frac{1}{2}\bar{{\cal F}_{2}}(\Box)\square+\frac{1}{3}\bar{{\cal F}_{3}}(\Box)\Box\right]\left(\partial_{\beta}\partial_{\alpha}h+\eta_{\alpha\beta}\partial_{\mu}\partial_{\nu}h^{\mu\nu}\right)\\
 & -\left[-1+2\bar{{\cal F}_{1}}(\Box)\Box+\frac{1}{2}\bar{{\cal F}_{2}}(\Box)\Box-\frac{1}{3}\bar{{\cal F}_{3}}(\Box)\Box\right]\eta_{\alpha\beta}\Box h\\
 & -\left[2\bar{{\cal F}_{1}}(\Box)\Box+\bar{{\cal F}_{2}}(\Box)\square+\frac{2}{3}\bar{{\cal F}_{3}}(\Box)\Box\right]\Box^{-1}\nabla_{\alpha}\nabla_{\beta}\partial_{\mu}\partial_{\nu}h^{\mu\nu}\,.
\end{aligned}
   \]
Here we have set $M_P^2\equiv 2$ for convenience. We can rewrite this as
\[
\label{Tababc}
\bl
T_{\alpha\beta} &=      -\biggl[{\bar a}(\Box)\square h_{\alpha\beta}+{\bar b}(\Box)\partial_{\sigma}\left(\partial_{\alpha}h_{\;\beta}^{\sigma}+\partial_{\beta}h_{\alpha}^{\;\sigma}\right)+{\bar c}(\Box)\left(\partial_{\beta}\partial_{\alpha}h+\eta_{\alpha\beta}\partial_{\mu}\partial_{\nu}h^{\mu\nu}\right)
\\&     +       {\bar d}(\Box)\eta_{\alpha\beta}\Box h+{\bar f}(\Box)\Box^{-1}\nabla_{\alpha}\nabla_{\beta}\partial_{\mu}\partial_{\nu}h^{\mu\nu}\biggr]\,,
 \el
 \]
where we have defined
\ba
\nn
{\bar a}(\Box)& \equiv & 1+\frac{1}{2}\bar{{\cal F}_{2}}(\Box)\square+\bar{{\cal F}_{3}}(\Box)\Box=-{\bar b}(\Box)\,, \\
\nn
{\bar c}(\Box ) &\equiv & 1-2\bar{{\cal F}_{1}}(\Box)\square-\frac{1}{2}\bar{{\cal F}_{2}}(\Box)\square+\frac{1}{3}\bar{{\cal F}_{3}}(\Box)\Box=-{\bar d}(\Box)\,, \\
{\bar f}(\Box)& \equiv & 2\bar{{\cal F}_{1}}(\Box)\Box+\bar{{\cal F}_{2}}(\Box)\square+\frac{2}{3}\bar{{\cal F}_{3}}(\Box)\Box\,, \label{abc}
 \ea
and have recovered the same constraints as in the UV~\footnote{We note that the forms of these constraints differ to those of Ref.~\cite{Biswas:2011ar}. This is due to different conventions, namely, in \cite{Biswas:2011ar}, the authors take the signature to be "mostly negative",  where as in Ref.~\cite{Biswas:2013cha}, we take the signature to be ``mostly positive" with $M_{p}^2=2$. Secondly, the presence of the Weyl tensor rather than the Riemann tensor in the action has an effect on the $\bar{{\cal F}_{3}}(\Box)$ terms. Having said this, when these convention changes are taken into account, we find that the above constraints are the same as those in \cite{Biswas:2011ar} and \cite{Biswas:2013cha} with the exception that we are now considering ${\bar {\cal F}}_i(\Box)=\sum^\infty_{n=1} f_{i_n}\Box^{-n}$ in the IR as opposed to ${\cal F}_i(\Box)=\sum^{\infty}_{n=0}f_{i_n}\Box^n$ in the UV.}~\cite{Biswas:2011ar, Biswas:2013cha}:
\ba
\nonumber
{\bar a}+{\bar b} & = & 0\,, \\
\nonumber
{\bar c}+{\bar d} & = & 0\,, \\
\label{constr}
{\bar b}+{\bar c}+{\bar f} & = & 0\,. 
 \ea
These equalities we found by explicit evaluation of the respective terms, can be understood as a consequence of the Bianchi identities. In the linearised limit, $\Box=\nabla_{\mu} \nabla^{\mu}=\partial^2$, and it suffices to take the partial derivative of \eqref{Tababc} as 
\be
\partial^{\beta}T_{\alpha\beta}  =-\lp {\bar a}+{\bar b}\rp\partial^{\sigma}\partial^{2}h_{\alpha\sigma}-\lp {\bar b}+{\bar c} + {\bar f}\rp \partial_{\alpha}\partial_{\mu}\partial_{\nu}h^{\mu\nu}-\lp {\bar c}+{\bar d} \rp \partial^{2}\partial_{\alpha}h\,.
\ee
This divergence should vanish identically, and when the coefficients of each independent term is zero due to (\ref{constr}), it does. It is this classical conservation structure of the theory that also sets the coefficients of the effective stress energy terms $\partial_{\beta}\partial_{\alpha}h$ and $\eta_{\alpha\beta}\partial_{\mu}\partial_{\nu}h^{\mu\nu}$ identical (denoted $-{\bar c}$ here) in the first place. 

We then close this section with a brief remark concerning massive gravity. The Fierz-Pauli term would have the form $M^2(h_{\mu\nu}-\eta_{\mu\nu})$. We can now indeed recover such a term, without resorting to Lorenz violation or additional metrics, with an action specified by an arbitrary $\bar{{\cal F}_{3}}$ by setting $\bar{{\cal F}_{1}} = M^2\Box^{-2} + 2\bar{{\cal F}_{3}}/3$ and $\bar{{\cal F}_{2}}=-2M^2\Box^{-2}-2\bar{{\cal F}_{3}}$. However, there will inevitably then also appear additional terms in the stress energy tensor, due to (\ref{constr}), and thus the linearised theory doesn't quite coincide with the pure Fierz-Pauli theory.

%\subsubsection*{Bianchi identities}
%It is worth noting that the constraints \eqref{constr} are equivalent to the Bianchi identities. For completeness we will sketch the proof more rigorously here. Taking the partial derivative of the linearised equation of motion given in \eqref{lineom} we obtain
%\[
%\begin{aligned}\partial^{\beta}T_{\alpha\beta} & =-\left[1+\frac{1}{2}\bar{{\cal F}_{2}}(\Box)\square+\bar{{\cal F}_{3}}(\Box)\Box-1-\frac{1}{2}\bar{{\cal F}_{2}}(\Box)\Box-\bar{{\cal F}_{3}}(\Box)\Box\right]\partial^{\sigma}\partial^{2}h_{\alpha\sigma}\\
% & -\biggl[-1-\frac{1}{2}\bar{{\cal F}_{2}}(\Box)\Box-\bar{{\cal F}_{3}}(\Box)\Box+2\bar{{\cal F}_{1}}(\Box)\Box+\bar{{\cal F}_{2}}(\Box)\square+\frac{2}{3}\bar{{\cal F}_{3}}(\Box)\Box\\
% & +1-2\bar{{\cal F}_{1}}(\Box)\square-\frac{1}{2}\bar{{\cal F}_{2}}(\Box)\square+\frac{1}{3}\bar{{\cal F}_{3}}(\Box)\Box\biggr]\partial_{\alpha}\partial_{\mu}\partial_{\nu}h^{\mu\nu}\\
% & -\biggl[1-2\bar{{\cal F}_{1}}(\Box)\square-\frac{1}{2}\bar{{\cal F}_{2}}(\Box)\square+\frac{1}{3}\bar{{\cal F}_{3}}(\Box)\Box\\
% & -1+2\bar{{\cal F}_{1}}(\Box)\Box+\frac{1}{2}\bar{{\cal F}_{2}}(\Box)\Box-\frac{1}{3}\bar{{\cal F}_{3}}(\Box)\Box\biggr]\partial^{2}\partial_{\alpha}h\,,
%\end{aligned}
 %\]
%where we have noted that in the weak field limit $\Box=\nabla_{\mu} \nabla^{\mu}=\partial^2$, we find after some simple algebra, all terms cancel
%\[
%\begin{aligned}\partial^{\beta}T_{\alpha\beta} & %=-[0]\partial^{\sigma}\partial^{2}h_{\alpha\sigma}-[0]\partial_{\alpha}\partial_{\mu}\partial_{\nu}h^{\mu\nu}-[0]\partial^{2}\partial_{\alpha}h=0\,.\end{aligned}
% \]

\subsection{Newtonian Potentials}

We wish to compute the Newtonian potentials. In order to do so, we consider the weak field (i.e. $h^2 \approx 0$) \emph{static} (i.e. $\Box \approx \nabla^2$) limit.
The trace and the $00$-component of the field equation \eqref{Tababc} are
\ba
-\rho & = &\left({\bar a}(\Box)-3{\bar c}(\Box)\right)\left(\square h-\partial_{\mu}\partial_{\nu}h^{\mu\nu}\right) \label{n1} \\
\rho  & = & {\bar a}(\Box)h_{00}+{\bar c}(\Box)\left(\Box h-\partial_{\mu}\partial_{\nu}h^{\mu\nu}\right) \label{n2}
 \ea
where we have assumed negligible pressures for $T_{\beta}^{\alpha}=diag(\rho,p,p,p)$, so that $tr(T_{\alpha\beta})=-\rho+3p\approx-\rho$ and $T_{00}=\rho$. We then impose the spherically symmetric metric
\[
ds^{2}=-(1+2\Phi)dt^{2}+(1-2\Psi)dr^{2}\,,
 \]
and note 
\[
h_{00}=-2\Phi,\quad h_{ij}=-2\Psi\eta_{ij}\,,
 \]
so that the pair of equations (\ref{n1},\ref{n2}) becomes
\ba
\nonumber
-\rho & = & 2\left({\bar a}-3{\bar c}\right)\left(\nabla^{2}\Phi-2\nabla^{2}\Psi\right)\,, \\
\label{rhoac}
\rho & = & 2\left({\bar c}-{\bar a}\right)\nabla^{2}\Phi-4{\bar c}(\Box)\nabla^{2}\Psi\,.
 \ea
%Therefore
%\[
%-4{\bar a}(\Box)\nabla^{2}\Phi=-4{\bar a}(\Box)\nabla^{2}\Psi=m\delta^{3}(\vec{r})
 %\]
Solving for $\Phi$, we can then, upon performing a Fourier transform and restoring the $M_p^2=1/(8\pi G)$, express the Newtonian potential as the integral %\textcolor{red}{Check the prefactors of the first equation}
\[
\label{Newtpot}
\Phi(r) = -\frac{m}{\lp 4\pi\rp^3 M_{p}^{2}}\int_{-\infty}^{\infty}d^{3}p\frac{e^{i\vec{p}\vec{r}}\lp \ab - 2\cb \rp}{2p^{2}\bar{a}\lp \ab-3\cb \rp}=-\frac{m}{\pi^2 M_{p}^{2} r}\int^{\infty}_{0}\frac{dp}{p}\frac{\sin(pr)\lp \ab-2\cb\rp}{2\bar{a}\lp \ab-3\cb\rp}\,,
 \]
where\footnote{However, we should point out a subtlety that though there is no ambiguity in the case of derivative operators, but we have identically that $\Box e^{ipr} = -p^2  e^{ipr}$, in the case of inverse derivative operators  $\Box^{-1} e^{ipr} = -p^{-2}  e^{ipr}$ implies a choice of boundary conditions for the operator $1/\Box$. These boundary conditions should be understood as specification of the operator and thus a property of the theory itself rather than free parameters for each solution. The boundary conditions adopted here amount to setting the homogeneous solution of the flat-space Green functions to zero, as seems most reasonable in this case. -A constant associated to the homogeneous modes was tuned in the screening mechanism of \cite{Nojiri:2010pw, Zhang:2011uv} to cancel the cosmological constant in cosmological background. It is unclear if such a prescription for the operator 
 $\Box^{-1}$ would be viable in other backgrounds.}
 $\ab=\ab(-p^2)$, $\cb=\cb(-p^2)$ and $m$ is the mass of the test particle. Similarly we get for $\Psi$ 
\be \label{Newtpot2}
\Psi(r) = \frac{m}{ \pi^2 M_{p}^{2} r}\int^{\infty}_{0}\frac{dp}{p}\frac{\sin(pr)\cb}{2\bar{a}\lp \ab-3\cb\rp}\,.
\ee
Thus we have now a complete algorithm to determine the Newtonian potentials of an arbitrary metric theory of gravity: given any form of local or non-local action, one may readily expand it up to quadratic order in the curvature, read off the functions $\ab$ and $\cb$ and obtain the post-Newtonian potentials by performing the above two integrals (\ref{Newtpot}, \ref{Newtpot}). 

We immediately see that, generally speaking, these will differ from each other, 
%$\Phi+\Psi \neq 0$. 
We describe their ratio using the Eddington parameter $\gamma$, which is defined as
\be
\gamma \equiv -\frac{\Psi}{\Phi}\,,
\ee
and is constrained by the Cassini tracking experiment to have the following upper bound $|\gamma -1|\leq 10^{-5}$ \cite{Will:2005va} and therefore the discrepancy can provide useful constraints on generic non-local models. However, we also notice that in the class of theories with $\ab=\cb$, i.e. $\bar{f}=0$, the Newtonian potentials will be identically the same, thus $\gamma=1$, but the potential can still deviate from the $1/r$ behaviour at large distances. This is in complete accordance with the results of \cite{Biswas:2011ar,Biswas:2013kla}, where the $a=c$ class of theories was found to introduce no new degrees of freedom, since the additional scalar contribution in the propagator disappears at the limit $a=c$, while the function $a$ still modulates the usual graviton propagator. In the case of non-analytical (inverse powers of the d'Alembertian) functions, however, the propagator may be an ill-defined object,
the intuition is retained here that the special $\bar{f}=0$ class of theories is devoid of an extra scalar and thus features only one independent Newtonian potential in the weak field
limit. To recapitulate, we have two classes of theories:
\begin{itemize}
\item $\bar{f} \neq 0$, i.e. $\ab \neq \cb$. In this case, we have an extra scalar degree of freedom and the two gravitational potentials are not independent, i.e. $\ga \neq 1$.
\item $\bar{f} = 0$, i.e. $\ab = \cb$. In this case, there are no additional modes and thus $\Psi=\Phi$ i.e. $\ga = 1$.
\end{itemize}
In the following section, we will consider explicit examples from both classes of theories.

\section{Examples} \label{sec:examples}

Armed with the machinery to study generic metric theories, we illustrate its power by applying it to several non-local models found in the literature, 
%The two first examples are known not to be viable as dark energy alternatives because of their failure to produce the observed large-scale structure. 
before proposing a new model featuring the exponential of the inverse d'Alembertian operator.

\subsection{The $Rf(R/\Box)$ Model}

The non-local model proposed by Deser and Woodard \cite{Deser:2007jk} is defined by the action
\be
S = \frac{M_p^2}{2}\int d^4 x \sqrt{-g} R\lb 1 + f(R/\Box)\rb\,.  
\ee
As  mentioned in the introduction, this model has been ruled out as an alternative to dark energy due to its impact on the structure formation \cite{Dodelson:2013sma}. It is, however, instructive to consider it first as perhaps the simplest example of a non-local modification of gravity in the infrared. Indeed, the $\gamma$ parameter turns out to be simply a constant up to the first order in the post-Newtonian expansion. For this order we only need the quadratic term
\be
S \approx \frac{M_p^2}{2}\int d^4 x \sqrt{-g} R\lp 1 + \frac{f'(0)}{\Box}R + \dots \rp\,.  
\ee
We may read off from the action that ${\cal F}_{1}=f'(0)/\Box$ and ${\cal F}_{2}={\cal F}_{3}=0$ as well as the functions $\ab=0$ and $\cb=2f'(0)$, upon referring back to (\ref{abc}). Solving the integral in (\ref{Newtpot},\ref{Newtpot2}) for these values of $\ab$ and $\cb$, we obtain
\be
\Phi = \frac{m\lp 1+8f'(0)\rp}{8\pi M_p^2 r \lp 1+6f'(0)\rp}\,, \quad \Psi = -\frac{m\lp 1+4f'(0)\rp}{8\pi M_p^2 r \lp 1+6f'(0)\rp}\,.
\ee
Newton's constant is thus shifted - an occurrence which can be nullified with a redefinition. To first order, we obtain $|\gamma-1| = 4f'(0)$. This agrees with an earlier result given in \cite{Koivisto:2008dh} and derived using a scalar field formulation of the theory  (see appendix \ref{sec:scalar} for such a treatment of ${\cal F}_{1}(\Box)$ theories). One can then test the validity of non-local models by substituting the prescribed value of $f'(0)$. For example, \cite{Park:2012cp} gives $f(X)$ as follows
\[
f(X)=0.245\biggl[\tanh(0.350Y+0.032Y^{2}+0.003Y^{3})-1\biggr],\qquad\mbox{with }Y\equiv X+16.5
 \]
 from which we deduce that 
 \[
 |\gamma-1|=7.77628\times10^{-24}
 \] 
 which is well within the Cassini bound. Conversely, for models of the type
 \[
 f(\phi)=f_{0}e^{\alpha\phi}
 \]
 as in \cite{Jhingan:2008ym}, with $\alpha>0.17$ and $f_0=\frac{\kappa^2 \rho_0}{3H_0^2(1+3w)}=\frac{\Omega_{0}}{(1+3w)}$ which is of order unity, we find that the best-fit values of the model are in disagreement with our constraints. In fact, $f_0$ must be less than $1.47\times 10^{-5}$ to be contained within the limits. 
\subsection{Non-Local Massive Gravity} 
\label{sec:MG}
%Consider the Lagrangian
%\[
%{\cal L}=\sqrt{-g}\biggl[\frac{2}{\kappa^{2}}R+R\alpha(\Box)R-R_{\mu\nu}\alpha(\Box)R^{\mu\nu}\biggr]
 %\]
%with 
%\[
%\alpha(\Box)=2\biggl(\frac{(\Box-m^{2})e^{H(-\Box_{M})}-\Box}{\kappa^{2}\Box^{2}}\biggr)
%\]
%where $H(-\Box_{M})$ is an entire function dependent on the operator $-\frac{\Box}{M^{2}}$
 %and $M$  is an ultraviolet mass scale and thus in our current (IR) treatment, we can set $H(-\Box_{M})$
%to zero. 
 As explained in section \ref{sec:limit}, in the context of non-local theories, the graviton may acquire mass without the introduction of an external reference metric \cite{Jaccard:2013gla,Modesto:2013jea}. The IR part of an action that has been proposed for this purpose reads
\[ \label{massive}
S=\frac{M_p^2}{2}\int d^4x {\sqrt{-g}}\biggl[R+R\left(\frac{M^{2}}{\Box^{2}}\right)R-2R_{\mu\nu}\left(\frac{M^{2}}{\Box^{2}}\right)R^{\mu\nu}\biggr]\,.
\]
As we will see shortly, we can now interpret the IR scale $M$ as the mass of the graviton. From the action, we read off
\[
{\cal \bar{F}}_{1}(\Box)\Box=\frac{M^{2}}{\Box},\qquad{\cal \bar{F}}_{2}(\Box)\Box=-2\frac{M^{2}}{\Box},\qquad\bar{{\cal F}}_{3}(\Box)=0\,,
 \]
so that
\[
\bar{a}(\Box)=\frac{\Box-M^{2}}{\Box}=-\bar{b}(\Box)=\bar{c}(\Box)=-\bar{d}(\Box)\,,
 \]
i.e.
\[
\bar{a}(-p^{2})=\frac{p^{2}+M^{2}}{p^{2}}\,.
 \]
We observe that this model belongs to the class $\bar{f}=0$ where there are no additional degrees of freedom. To solve for the gravitational potential (\ref{Newtpot}), we need to integrate ${\cal I}=\int_{0}^{\infty}\frac{\sin(pr)}{p\bar{a}(-p^{2})}dp$, that is
\[
{\cal I}=\int_{0}^{\infty}\frac{p\sin(pr)}{p^{2}+M^{2}}dp=\int_{0}^{\infty}\frac{pe^{ipr}-pe^{-ipr}}{2i(p+iM)(p-iM)}dp\,,
 \]
where $M$ represents the mass of the graviton. We observe that there are poles at $p=iM$
  on the upper half plane and $p=-iM$
  on the lower half plane. We will take each pole separately and use the general contour integral formula
\[
\label{contform}
\oint_{C}\frac{f(z)}{z-z_{0}}dz=2\pi if(z_{0})\,.
 \]
Pole $p=iM$
  is on the upper half plane so we only consider the $e^{ipr}$
  portion of $\sin(pr)$
  which encircles the pole, so that
\[
{\cal I}=\oint_{m}\frac{pe^{ipr}/2i(p+iM)}{(p-iM)}dp\,.
 \]
which we consider of the form
\[
{\cal I}=\oint_{m}\frac{f(p)}{p-p_{0}}dp=2\pi if(p_{0})\,.
 \]
with $p_{0}=iM$  and $f(p)=pe^{ipr}/2i(p+iM)$, so that
\[
{\cal I}=2\pi if(iM)=\frac{\pi}{2}e^{-Mr}\,.
 \]
Similarly for the pole $p=-iM$
\[
{\cal I}=\oint_{-m}\frac{-pe^{-ipr}/2i(p-iM)}{(p+iM)}dp=-2\pi if(-iM)=-2i\pi\left(\frac{iMe^{-Mr}}{4M}\right)=\frac{\pi}{2}e^{-Mr}\,,
 \]
where, in this case we are moving into the lower half plane so that
\[
{\cal I}=\oint_{-m}\frac{f(p)}{p-p_{0}}dp=-2\pi if(p_{0})\,.
 \]
Hence
\[
{\cal I}=\frac{\pi}{2}e^{-Mr}\,.
 \]
%From \eqref{Newtpot}, we can then write the Newtonian potential as
%\[
%\Phi(r)=\frac{m}{4\pi^{2}M_{p}^{2} r}\int dp\frac{\sin(pr)}{p\; a(-p^{2})}=\frac{4GM}{\pi r}\int dp\frac{Sin(pr)}{p\; a(-p^{2})}
 %\]
%with $\kappa=8\pi G$
%$M_{P}=\sqrt{\frac{\hbar c}{8\pi G}}$
%and $\hbar=c=1$.
Solving the integral \eqref{Newtpot}, we find
\[ \label{massivephi}
\Phi(r)	=	\frac{m}{4\pi^{2} M_{p}^{2} r}\int^{\infty}_{0} dp\frac{p\sin(pr)}{p^{2}+M^{2}}=\frac{m}{4\pi^{2}rM_{p}^{2}}\left(\frac{\pi}{2}e^{-Mr}\right)=\frac{m}{8\pi rM_{p}^{2}}e^{-Mr}\,.
 \]
We thus obtain precisely the expected Yukawa-type correction for massive gravity. We plot the solution (\ref{massivephi}) in Figure \ref{Phiplot}.

\subsection{The $R\Box^{-2}R$ Model and Generalisations}

A variation of the previous model (\ref{massive}) where the tensorial piece is omitted was studied in Refs. \cite{Maggiore:2014sia,Dirian:2014ara}, where the instability arising from tensorial non-localities is avoided \cite{Foffa:2013vma,Ferreira:2013tqn}. However, as we have learned, an additional scalar mode will appear in flat space. The action to consider 
is\footnote{Here we chose the opposite sign for the $M^2$-term from \cite{Maggiore:2014sia}. In the case of a scalar field at least, that sign choice would correspond to tachyonic mass-squared, which, however, is the choice that has been shown to lead to interesting cosmology \cite{Maggiore:2014sia,Dirian:2014ara}. The Newtonian limit has been calculated for that case in Ref. \cite{Kehagias:2014sda}, expectedly with different results from what we obtain here. Setting $M\rightarrow iM$, one obtains oscillating $\cos{Mr}$-type corrections instead of the exponential $e^{-Mr}$ we find here and in (\ref{massivephi}).}
\be \label{massive2}
S = \frac{M_p^2}{2}\int d^4x \sqrt{-g}R\lp 1 + \frac{1}{3}\frac{ M^2}{\Box^2} R\rp\,.
\ee
Thus, ${\cal \bar{F}}_{1}(\Box)=M^2/3\Box^2$ and ${\cal \bar{F}}_{2}(\Box)={\cal \bar{F}}_{3}(\Box)=0$. We proceed analogously to the previous two cases, further details can be found in Appendix \ref{sec:appCI}. % \textcolor{red}{Aindriu you can add some details if you prefer}. 
 As before, the integrals in (\ref{Newtpot}, \ref{Newtpot2}) can be completed by calculating their residuals, resulting in the following:
\be
\Phi(r)=\frac{\left(4-e^{-Mr}\right)m}{24M_{P}^{2}\pi r},\qquad\Psi(r)=\frac{-(2+e^{-Mr})m}{24M_{P}^{2}\pi r}\,.
 \ee
Thus, the gravitational potentials differ from each other and display the usual $1/r$ behaviour at distances $r \gtrsim r_0\sim 1/M$, as expected. 

Here, one may use the Cassini bound to put limits on the mass of the graviton in order to verify if it lies on the dark energy scale. We remind the reader that in order to be a suitable dark energy candidate, the graviton must be of  the order of the present Hubble parameter $\sim 10^{-33}$. Taking
\[
|\gamma-1|=\frac{2-2e^{-Mr}}{4-e^{-Mr}}\leq1\times10^{-5},
 \]
 we then expand the exponential terms to first order, whilst setting $r=5.06773\times10^{18}eV^{-1}$ or 1 billion kilometres as in the original Cassini experiment \cite{Will:2005va}, from which we find the upper bound on the mass of the graviton in a theory of this type to be
 \[
 M\leq2.95992\times10^{-24}eV.
 \]
Subsequently, with a small enough mass, this particular model of massive gravity is within the permissible limits of being a suitable candidate for dark energy.

The asymptotic value of the Eddington parameter $\gamma \rightarrow 1/2$ coincides with that predicted for Solar System measurements in fourth order (i.e. local) metric $f(R)$ theories \cite{Chiba:2003ir} (for a unified analysis covering also e.g. the Palatini and non-minimally coupled $f(R)$ theories, see \cite{Koivisto:2011tp}). The Newtonian potentials behave contrary to those in $f(R)$ models: near to the source $\ga=1/2$ and at large distances $\ga \rightarrow 1$. This can readily be seen by plugging a constant ${\cal \bar{F}}_{1}=f''(0)$ into our expressions, for instance ${\cal \bar{F}}_{1}=r^2_0/3$:
\be
\Phi_{f(R)}(r) = \frac{\lp 3+e^{-\frac{r}{r_0}} \rp m}{24M_p^2 \pi r}\,, \quad \Psi_{f(R)}(r) = -\frac{\lp 3-e^{-\frac{r}{r_0}} \rp m}{24M_p^2 \pi r}\,.
\ee
It is nontrivial that non-local models of the type (\ref{massive2}) exhibit the opposite behaviour with respect to the fourth order local models. Were this not the case, however, the former would of course be immediately ruled out.

\begin{figure}
\begin{center}
\includegraphics[width=10cm]{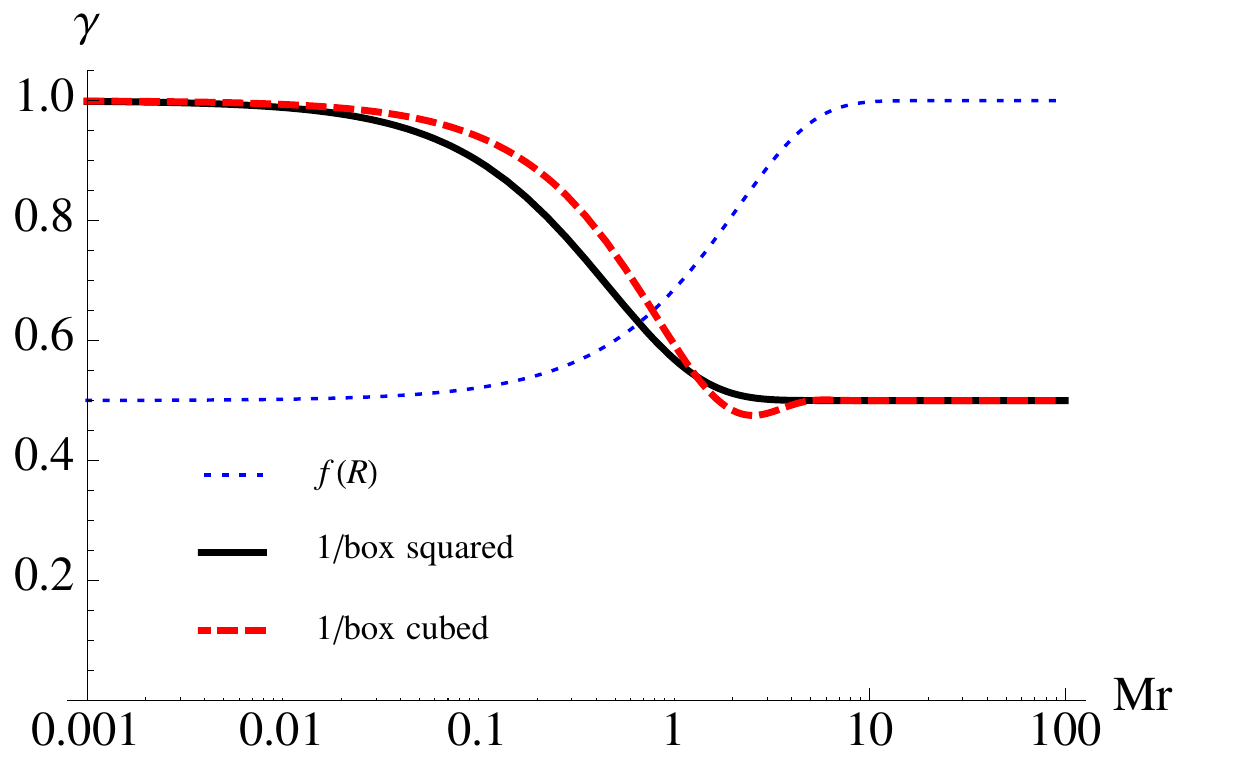}
\caption{\label{gammaplot}
The Eddington parameter as a function of the distance from the source in $R^2$ models. The thick black line is for the model (\ref{massive2}), the thin blue dotted line is for the local $f(R)$ gravity and the dashed red line for a $RM^4\Box^{-3}R$ model. For higher powers of $1/\Box$, the behaviour is qualitatively the same.}
\end{center}
\end{figure}

We have checked that by considering higher powers of the inverse box operator, for example a $RM^4\Box^{-3}R$ model, one obtains qualitatively similar behaviour with exponential modification terms, further modulated by oscillatory functions. We illustrate this in figure \ref{gammaplot}, where we plot the $\gamma$ for both the $RM^2\Box^{-2}R$ and $RM^4\Box^{-3}R$ models.

\subsection{Degravitation with ${\cal \bar F}_{i}(\Box)$ and $\bar a =\bar c\neq 0,~\bar f=0$ }

\begin{figure}
\begin{center}
\includegraphics[width=10cm]{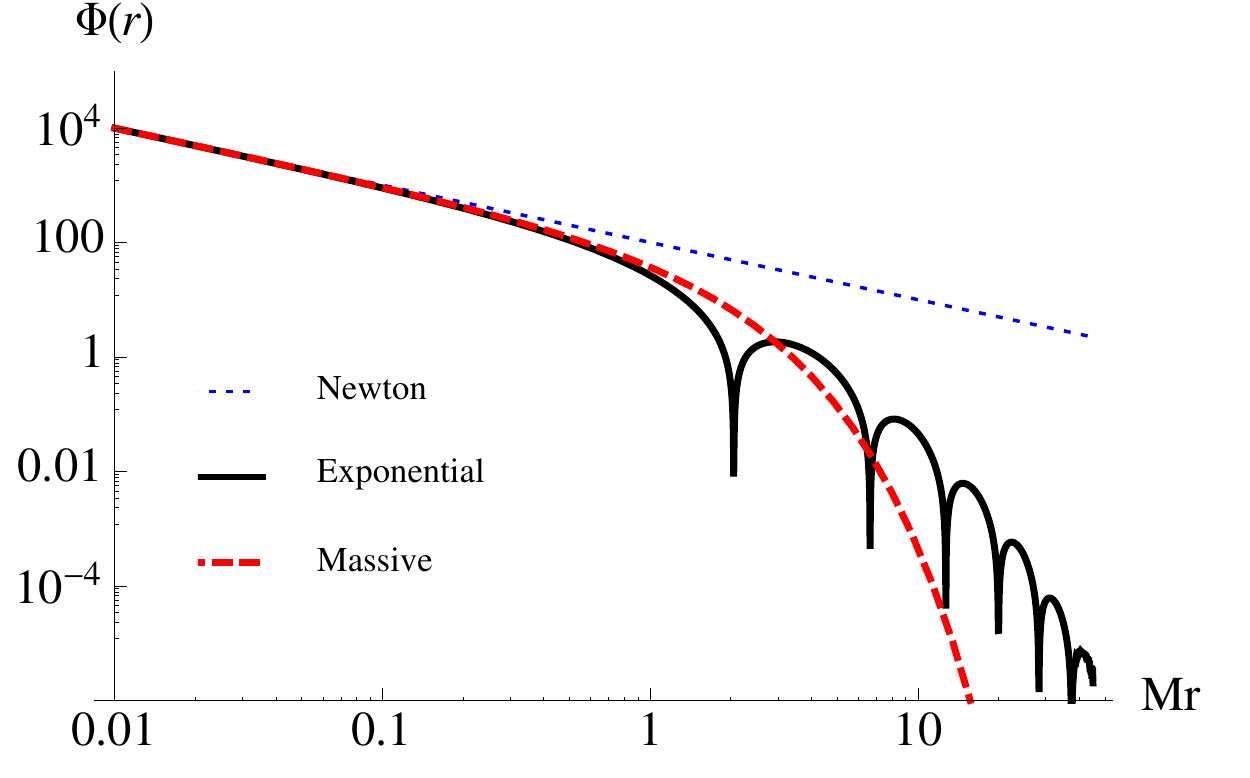}
\caption{\label{Phiplot}
The suppression of the gravitational potential in the exponential model. The thick black line is the modulus of the potential (\ref{solveint}) as a function of the radius. Few first oscillations are visible in the figure as the potential is suppressed with respect to the Newtonian $1/r$ behaviour depicted by the thin blue dotted line. For comparison, we also show the pure Yukawa suppression in massive gravity (\ref{massivephi}) as the dashed red line.}
\end{center}
\end{figure}

Instead of keeping $\Box^{-1}$ order by order,  in this section we wish to keep all orders in ${\cal \bar F}_{i}(\Box) =\sum_{n=1}^{\infty}f_{i_n}\Box^{-n}$. 
We study the scenario when $\bar a =\bar c\neq 0,~\bar f=0$. One particular choice would be to consider the function ${\bar a}(\Box)={\bar c}(\Box)=e^{-{M^2}/{\Box}}$, 
see Appendix~\ref{sec:scalar}. This, as other examples considered here, is a non-analytic function. It is however qualitatively different as it involves an infinite series of inverse derivative operators, featuring `'double non-locality'' in this sense. The would-be propagator has an essential singularity at the origin. The zero-mode of course corresponds to the cosmological constant, and it remains to be seen what the presence of an essential singularity implies for degravitation. Here our aim is simply to check that the model behaves reasonably at its Newtonian limit. 
We obtain:
%Note that in (5.13) the boundary of integral is changed since there has to be a source (can we elaborate on this?). 
\begin{equation}
\label{solveint}
\Phi(r)=\frac{4\pi m}{M_P^2}\biggl[\frac{2 \pi ^2 \text{sgn}(r) \, _0F_2\left(;\frac{1}{2},1;\frac{M^2 r^2}{4}\right)}{r}-4 M \pi ^{3/2} \, _0F_2\left(;\frac{3}{2},\frac{3}{2};\frac{M^2 r^2}{4}\right)\biggr]\,.
\end{equation}
where \(_pF_q\) is the hypergeometric function with \(p=0\) and \(q=2\). %/ One can then plot \eqref{solveint} as follows, where for convenience we have set the constants $m, M$ and $M_P$ to 1. 
We plot the $r$-dependence of the potential in figure \ref{Phiplot}. As the modifications become significant for large $r$, the potential begins to oscillate and changes its sign rapidly. The latter feature, which was not present in the previus examples, could have interesting effects at astrophysical and cosmological scales when $r\gtrsim 1/M$. Note that the Newtonian potential drops rapidly at large distances, exhibiting the degravitation of the Newtonian potential in the far IR. 

\section{Concluding Remarks} \label{sec:conclusions}

In this treatment, we have analysed general, non-local infrared modifications of gravity. Specifically, we considered cases where the gravitational action is quadratic in the curvature and the infrared non-localities can be presented as (possible infinite) power series of the inverse d'Alembertian operator. This suffices in order to study the leading order weak-field limit of {\it any} metric theory of gravity, since for that one needs the action only up to quadratic order in the curvature. 

The full field equations were presented in the form Eqs.~(\ref{eomir}, \ref{tensors}) and they were cross-checked by explicit computation of the Bianchi identities. We then considered the weak field limit of the theories in order to calculate the gravitational potentials from which one could extract the post-Newtonian corrections. We identified a specific class of theories  in which there is only one independent gravitational potential at the weak-field limit. The models within this class involve both scalar and tensor non-localities, specified by a single free function. Non-local theories of the form $Rf(\Box^{-n}R)$ do not include tensorial non-localities and thus feature an additional scalar degree of freedom. Therefore they predict a non-trivial Eddington parameter $\gamma$ that describes the ratio of the two independent gravitational potentials. Regarding this class of theories, we can summarise our findings as follows:
\begin{itemize}
\item n=0: In the fourth order (local) $f(R)$ models, near the source we have $\gamma =1/2$, whereas at large radii $\ga\rightarrow 1$. 
\item n=1: In this class of models the $\gamma$ is a constant.
\item n=2: In these variations of massive gravity (without the discontinuity), near the sources $\gamma =1$ and at large radii $\ga\rightarrow 1/2$.
\item n>2: With higher order non-locality, the asymptotic behaviours remain the same as in the previous case, but GR modifications of type $e^{-M r}$ are modulated by oscillations of the type $\cos(Mr)$.
\end{itemize}
Furthermore, we checked that the non-local massive gravity action predicts the expected Yukawa-type correction to the Newtonial potential.

A new model was proposed, specified by the single function $\exp{\lp -{M^2}/{\Box} \rp}$. In this scenario, the Eddigton parameter is identically unit and the Newtonian potential is, at distances larger than $r\sim 1/M$ oscillationg with an exponentially suppressed amplitude. We expect this model to exhibit degravitation, as gravity is very strongly suppressed in the far infrared modes. Some properties of this theory with infinite-order inverse derivative operators could be quite different when compared to theories with finite order inverse operators and it would be interesting to explore further. For example, one could investigate whether the growing modes associated to tensorial non-local terms in power-law models will be present in the exponential model. In the ultraviolet, it is indeed crucial to consider entire functions, i.e. infinite series of derivatives, in order to bestow asymptotic freedom without introducing any new dangerous degrees of freedom that would spoil unitarity. It remains to be seen whether we can free gravity from the weight of the vacuum in the furthermost infrared by not truncating the order of the operators without giving rise to new instabilities.

\acknowledgments
We would like to thank Tirthabir Biswas, Alexey Koshelev, Michele Maggiore, Spyridon Talaganis, Sergey Vernov and to both referees of \emph{CQG} for useful discussions. 
AM is supported by the Lancaster-Manchester-Sheffield Consortium for Fundamental Physics under STFC grant ST/J000418/1. AC is funded by STFC grant no ST/K50208X/1.

\appendix

\section{Scalar Presentation} \label{sec:scalar}

Consider the action
\begin{equation}
\label{motaction}
S=\int d^{4}x\sqrt{-g}\biggl(R+R\sum^{\infty}_{n=1}a_{n}\Box^{-n}R\biggr)\,.
\end{equation}
We can rewrite this action in the form of a higher derivative scalar-tensor action
\begin{equation}
\label{scatens}
S=\int d^{4}x\sqrt{-g}\biggl(\Phi^{-1} R+\sum^{\infty}_{n=1}a_{n}\psi\Box^{-n}\psi-\psi\Phi^{-1}+\psi\biggr)\,,
\end{equation}
where if we set \(\psi=R\), the action \eqref{motaction} is recovered. We now perform a conformal transformation, 
where we have defined the following:
\begin{equation}
\Phi=e^{\phi}\approx1+\phi +\mathcal{O}(\phi^{2})\quad \implies \quad 1/\Phi=1/e^{\phi}\approx1-\phi+\mathcal{O}(\phi^{2})\,.
\end{equation}
Subsequently, up to quadratic orders, we obtain the following
\begin{equation}
S=\int d^{4}x\sqrt{-\tilde{g}}\biggl(\tilde{R}-\frac{3}{2}\phi\tilde{\Box}\phi+\sum^{\infty}_{n=1}a_{n}\psi\tilde{\Box}^{-n}\psi+\psi\phi\biggr)\,.
\end{equation}
Varying the action for \(\phi\) and \(\psi\), gives
\begin{equation}
\label{ds1}
\frac{\delta S}{\delta\phi}=\psi -3 \Box \phi=0\qquad\implies\qquad\psi= 3 \Box \phi\,,
\end{equation}
\begin{equation}
\label{ds2}
\frac{\delta S}{\delta\psi}=2  \sum^{\infty}_{n=1}a_{n}\Box^{-n}\psi+\phi=0\qquad\implies\qquad\phi=-2
\sum^{\infty}_{n=1}a_{n}\Box^{-n}\psi\,.
\end{equation}
Substituting \(\psi\) from \eqref{ds1} into \eqref{ds2} gives
\begin{equation}
-6\sum^{\infty}_{n=1}a_{n}\frac{\Box}{
\Box^{n}}\phi=-6\sum^{\infty}_{n=1}a_{n}\frac{1}{\Box^{n-1}}\phi\equiv\tilde{\Gamma}(\Box)\phi = 0\,.
\end{equation}
Here, we may identify $\tilde{\Gamma}(\Box)$ as something analogous to the propagator.
We would like to consider an infrared model with the inverse exponential modulating the propagator. Thus, we write
%\begin{displaymath}
%\tilde{\gamma}(\omega)=-\omega^{-1}\Rightarrow\tilde\Gamma(\omega)=e^{-1/\omega}\,,
%\end{displaymath}
%which implies that
\begin{equation}
\label{einverse}
-6\sum^{\infty}_{n=1}a_{n}\frac{1}{\Box^{n-1}}\equiv e^{-M^2/\Box}\,.
\end{equation}
before deriving appropriate coefficient  
\begin{equation}
\label{coef}
a_{n}=-\frac{1}{6}\frac{(-1)^{n-1}M^{2(n-1)}}{(n-1)!}
\end{equation}
We note that \eqref{einverse} is an entire function and for the  coefficients above, we find
\begin{equation}
\label{entire}
R-\frac{1}{6}R\sum^{\infty}_{n=1}\frac{(-1)^{n-1}M^{2(n-1)}}{(n-1)!}\Box^{-n}R=R-\frac{1}{6}R\biggl(\frac{e^{-M^2/\Box}}{\Box}\biggr)R\,.
\end{equation}
%\textit{CHECK (2.12) PLEASE! ARE WE USING \(\bar{{\cal
%F}}(\Box)=(\frac{e^{-1/\Box}-1}{\Box}) \) OR \(\bar{{\cal
%F}}(\Box)=e^{-1/\Box}\)??}
This argument can then be generalised for higher orders as  appear in the action  \eqref{action}. 
\section{Variation of the Inverse D'Alembertian Expanded}
\label{sec:app2-1}
Following on from section \ref{2-1}, we give further details of the computation of $\delta(\Box^{-1})$. From the product rule, we have
\[
\label{deltaminus}
\delta(\Box^{-1})S=-\Box^{-1}\delta(\Box)\Box^{-1}S+\delta(\Box^{-1}\Box)\Box^{-1}S
 \]
We note now that though $\Box\Box^{-1}S=S$, in general we can not say that $\Box^{-1}\Box S=S$. However, there is a subtlety regarding the precise nature of the inverse D'Alembertian, which we shall describe briefly below. As discussed in \cite{Foffa:2013vma}, the inverse D'Alembertian may be expressed in terms of the Green's function as follows
\[
(\Box^{-1}j)(x)\equiv f_{hom}(x)+\int d^{d+1}y\sqrt{-g(y)}G(x,y)j(y),\qquad\Box_{x}G(x,y)=\frac{1}{\sqrt{-g(x)}}\delta^{(d+1)}(x-y)
 \]
where $f_{hom}$  is the homogenous solution, i.e. any solution satisfying $\Box f_{hom}(x)=0$. Subsequently, by setting $j\equiv\Box S$, we find
\[
(\Box^{-1}\Box S)(x)=f_{hom}(x)+S(x)
 \]
Next, we note that from $\delta(\Box\Box^{-1})S=0$, we have
\[
\delta(\Box)\Box^{-1}S=-\Box\delta(\Box^{-1})S
 \]
We also note that by acting on \eqref{deltaminus} from the left hand side with $\Box$,  we retrieve this identity if
\[
\Box\delta(\Box^{-1}\Box)\Box^{-1}S=0
 \]

Thus, we may take
\[
f_{hom}(x)=\delta(\Box^{-1}\Box)\Box^{-1}S(x)
 \]

as $f_{hom}(x)$  is defined as any function such that $\Box f_{hom}(x)=0$. Therefore
\[
\delta(\Box^{-1})S=-\Box^{-1}\delta(\Box)\Box^{-1}S+f_{hom}(x)
 \]

Substituting this into \eqref{delta-n}, we find
\[
\delta(\Box^{-n})S=-\sum_{m=0}^{n-1}\Box^{-m-1}\delta(\Box)\Box^{-n+m}S+\sum_{m=0}^{n-1}\Box^{-m}f_{hom}(x)\Box^{-n+m+1}S
 \]

For $\bar{{\cal F}}_{i}(\Box)=\sum_{n=1}^{\infty}f_{i_{-n}}\Box^{-n}$, , we then arrive at
\[
\delta\bar{{\cal F}}_{i}(\Box)S=-\sum_{n=1}^{\infty}f_{i_{-n}}\sum_{m=0}^{n-1}\Box^{-m-1}\delta(\Box)\Box^{-n+m}S+\sum_{n=1}^{\infty}f_{i_{-n}}\sum_{m=0}^{n-1}\Box^{-m}f_{hom}(x)\Box^{-n+m+1}S
 \]

Now, as discussed in \cite{Foffa:2013vma}, \cite{Dirian:2014xoa}, non-local operators such as the inverse D'alembertian can be expressed in both retarded and advanced Green's functions, the result of which is that the correlating equations of motion are acausal. In order to preserve causality we must assume that only retarded Green's functions are considered by imposing
\[
G(x,y)=0\qquad\mbox{unless }y\mbox{ is in the past light-cone of }x
 \]
Secondly, in order to ensure that the prescribed action is generally covariant and the energy-momentum tensor is conserved, we must impose the boundary conditions
\[
G(x,y)|_{x^{0}=t_{0}}=0,\qquad\partial_{0}G(x,y)|_{x^{0}=t_{0}}=0
 \]
\cite{Foffa:2013vma}, \cite{Dirian:2014xoa}, which demands that the non-local effects start at $t_{0}$. Taking all this into account, in order to achieve a generally covariant solution, we consider only solutions with vanishing homogenous solution, thus
\[
\delta\bar{{\cal F}}_{i}(\Box)S=-\sum_{n=1}^{\infty}f_{i_{-n}}\sum_{m=0}^{n-1}\Box^{-m-1}\delta(\Box)\Box^{-n+m}S
 \]
\section{Variational Terms}
\label{sec:var}
\[
\delta(\Box)R=-h_{\alpha\beta}R^{;\alpha;\beta}+\frac{1}{2}g^{\alpha\beta}R_{;\lambda}(h_{\alpha\beta})^{;\lambda}-R^{;\alpha}(h_{\alpha\beta})^{;\beta}
\]
 \[
\bl
\delta(\square)R_{\mu\nu}       &=      -h_{\alpha\beta}R_{\mu\nu}^{;\alpha;\beta}-(h_{\alpha\beta})^{;\beta}R_{\mu\nu}^{;\alpha}+\frac{1}{2}g^{\alpha\beta}(h_{\alpha\beta})^{;\sigma}R_{\mu\nu;\sigma}
                \\&-\frac{1}{2}\left[\square(h_{\alpha\beta})\delta_{(\mu}^{\beta}R_{\;\nu)}^{\alpha}-(h_{\alpha\beta})^{;\tau;\alpha}\delta_{(\mu}^{\beta}R_{\tau\nu)}+(h_{\alpha\beta})_{;(\mu}^{\;;\beta}R_{\;\nu)}^{\alpha}\right]
                \\&-R_{\;(\nu}^{\alpha;\beta}h_{\alpha\beta;\mu)}-\delta_{(\mu}^{\beta}R_{\;\nu)}^{\alpha;\lambda}h_{\alpha\beta;\lambda}+\delta_{(\mu}^{\beta}R_{\tau\nu)}^{;\alpha}h_{\alpha\beta}^{\;\;;\tau})
\el
\] 
\[
\bl
\delta(\Box)R_{\mu\nu\lambda\sigma}     &=      -h_{\alpha\beta}R_{\mu\nu\lambda\sigma}^{;\alpha;\beta}-(h_{\alpha\beta})^{;\beta}R_{\mu\nu\lambda\sigma}^{;\alpha}+\frac{1}{2}h{}^{;\tau}R_{\mu\nu\lambda\sigma;\tau}
                \\&-\frac{1}{2}\biggl[g^{\alpha\tau}(h_{\alpha\beta})_{;\mu}^{;\beta}R_{\tau\nu\lambda\sigma}+g^{\alpha\tau}(h_{\alpha\beta})_{;\nu}^{;\beta}R_{\mu\tau\lambda\sigma}+g^{\alpha\tau}(h_{\alpha\beta})_{;\lambda}^{;\beta}R_{\mu\nu\tau\sigma}+g^{\alpha\tau}(h_{\alpha\beta})_{;\sigma}^{;\beta}R_{\mu\nu\lambda\tau}\biggr]
                \\&-\left[g^{\alpha\tau}(h_{\alpha\beta})_{;\mu}R_{\tau\nu\lambda\sigma}^{;\beta}+g^{\alpha\tau}(h_{\alpha\beta})_{;\nu}R_{\mu\tau\lambda\sigma}^{;\beta}+g^{\alpha\tau}(h_{\alpha\beta})_{;\lambda}R_{\mu\nu\tau\sigma}^{;\beta}+g^{\alpha\tau}(h_{\alpha\beta})_{;\sigma}R_{\mu\nu\lambda\tau}^{;\beta}\right]
\el
 \]
\[
\delta R_{\mu\nu\lambda\sigma}=\frac{1}{2}[\delta_{\lambda}^{\alpha}\delta_{\nu}^{\beta}(h_{\alpha\beta})_{;\sigma;\mu}-\delta_{\lambda}^{\alpha}\delta_{\mu}^{\beta}(h_{\alpha\beta})_{;\sigma;\nu}+\delta_{\mu}^{\alpha}\delta_{\sigma}^{\beta}(h_{\alpha\beta})_{;\nu;\lambda}-\delta_{\sigma}^{\alpha}\delta_{\nu}^{\beta}(h_{\alpha\beta})_{;\mu;\lambda}]
 \]
\[
\delta R_{\mu\nu}=\frac{1}{2}[\delta_{\nu}^{\beta}(h_{\alpha\beta})_{;\mu}^{;\alpha}+\delta_{\mu}^{\beta}(h_{\alpha\beta})_{;\nu}^{;\alpha}-\delta_{\mu}^{\alpha}\delta_{\nu}^{\beta}\square(h_{\alpha\beta})-g^{\alpha\beta}(h_{\alpha\beta})_{;\mu;\nu}]
 \]
\[
C_{\mu\nu\lambda\sigma}\delta{\cal F}_{3}C^{\mu\nu\lambda\sigma}=\left(2R_{\mu\nu}{\cal F}_{3}C_{\mu\nu\lambda\sigma}+({\cal F}_{3}C_{\mu\nu\lambda\sigma})_{;\mu;\nu}\right)h_{\alpha\beta}\,.\qquad\mbox{with }C_{\;\nu\mu\lambda}^{\mu}=0
 \]
\[
\delta R=-h_{\alpha\beta}R^{\alpha\beta}+(h_{\alpha\beta})^{;\alpha;\beta}-g^{\alpha\beta}\square(h_{\alpha\beta})
 \]
\[
\delta\sqrt{-g}=\frac{1}{2}\sqrt{-g}g^{\alpha\beta}h_{\alpha\beta}
 \]
 \section{Bianchi Identities}
 Below, we sketch the proof for the Bianchi identities of the Ricci and Weyl tensor sections of the action \eqref{action}.
 \label{sec:Bianchi}
 \subsection{$S_2$}
\[
S_{2}=\int d^{4}x\sqrt{-g}\left(R_{\mu\nu}\bar{{\cal F}}_{2}(\Box)R^{\mu\nu}\right)
 \]

The equation of motion reads
\[
\begin{aligned}T_{\alpha\beta}^{2} & =2R_{\alpha\sigma}\bar{{\cal F}_{2}}(\Box)R_{\;\beta}^{\sigma}-\frac{1}{2}g_{\alpha\beta}R_{\nu}^{\mu}\bar{{\cal F}_{2}}(\Box)R_{\mu}^{\nu}-2\nabla_{\sigma}\nabla_{\beta}\bar{{\cal F}_{2}}(\Box)R_{\alpha}^{\;\sigma}\\
 & +\square\bar{{\cal F}_{2}}(\Box)R_{\alpha\beta}+g_{\alpha\beta}\nabla_{\mu}\nabla_{\nu}\bar{{\cal F}_{2}}(\Box)R^{\mu\nu}+\Theta_{\alpha\beta}^{2}-\frac{1}{2}g_{\alpha\beta}\left(\Theta_{\sigma}^{2\sigma}+\bar{\Theta}^{2}\right)+2{\cal E}_{\alpha\beta}^{2}\,.
\end{aligned}
 \]

Taking the covariant derivative, we find
\[
\begin{aligned}\nabla^{\beta}T_{\alpha\beta}^{2} & =2\nabla^{\lambda}R_{\alpha\sigma}\bar{{\cal F}_{2}}(\Box)R_{\;\lambda}^{\sigma}+2R_{\alpha\sigma}\nabla^{\lambda}\bar{{\cal F}_{2}}(\Box)R_{\;\lambda}^{\sigma}-\frac{1}{2}\nabla_{\alpha}R_{\nu}^{\mu}\bar{{\cal F}_{2}}(\Box)R_{\mu}^{\nu}\,,
\\&
-\frac{1}{2}R_{\nu}^{\mu}\nabla_{\alpha}\bar{{\cal F}_{2}}(\Box)R_{\mu}^{\nu}-2\nabla^{\lambda}\nabla_{\sigma}\nabla_{\lambda}\bar{{\cal F}_{2}}(\Box)R_{\alpha}^{\;\sigma}+\nabla^{\lambda}\square\bar{{\cal F}_{2}}(\Box)R_{\alpha\lambda}+\nabla_{\alpha}\nabla_{\mu}\nabla_{\nu}\bar{{\cal F}_{2}}(\Box)R^{\mu\nu}\,,
\\&
+\nabla^{\sigma}\Theta_{\alpha\sigma}^{2}-\frac{1}{2}\nabla_{\alpha}\Theta_{\sigma}^{2\sigma}-\frac{1}{2}\nabla_{\alpha}\bar{\Theta}^{2}+2\nabla^{\sigma}{\cal E}_{\alpha\sigma}^{2}\,,
\end{aligned}
 \]
Using \eqref{tensors}, we solve for the following
\[
\nn\nabla^{\sigma}\Theta_{\alpha\sigma}^{2}=\sum_{n=1}^{\infty}f_{2_{-n}}\sum_{l=0}^{n-1}\left[\nabla^{\sigma}\nabla_{\alpha}R_{\nu}^{\mu(-l-1)}\nabla_{\sigma}R_{\mu}^{\nu(l-n)}+\nabla_{\alpha}R_{\nu}^{\mu(-l-1)}R_{\mu}^{\nu(-n+l+1)}\right]\,,
   \]
\[
\nn
-\frac{1}{2}\nabla_{\alpha}\Theta_{\sigma}^{2\sigma}=\sum_{n=1}^{\infty}f_{2_{-n}}\sum_{l=0}^{n-1}\left[-\nabla_{\alpha}\nabla_{\sigma}R_{\nu}^{\mu(-l-1)}\nabla^{\sigma}R_{\mu}^{\nu(l-n)}\right]\,,
   \]
\[
\nn
-\frac{1}{2}\nabla_{\alpha}\bar{\Theta}^{2}=\sum_{n=1}^{\infty}f_{2_{-n}}\sum_{l=0}^{n-1}\left[-R_{\mu}^{\nu(-l-1)}\nabla_{\alpha}R_{\nu}^{\mu(-n+l+1)}\right]\,,
   \]
\[
\nabla^{\lambda}{\cal E}_{\alpha\lambda}^{2}=\sum_{n=1}^{\infty}f_{2_{-n}}\sum_{l=0}^{n-1}\biggl[[\nabla_{\nu},\nabla^{\lambda}]R_{\lambda}^{\;\sigma(l-n)}\nabla_{\alpha}R_{\;\sigma}^{\nu(-l-1)}+[\nabla_{\nu},\nabla^{\lambda}]\nabla_{\alpha}R_{\;\sigma}^{\nu(l-n)}R_{\lambda}^{\;\sigma(-l-1)}\biggr]\,.
   \]
Finally using the following general formula
\[
\label{covcom}
\bl
{[}\nabla_{\rho},\nabla_{\sigma}{]}X_{\;\;\;\;\;\nu_{1}...\nu_{l}}^{\mu_{1}...\mu_{k}}	&=	R_{\;\lambda\rho\sigma}^{\mu_{1}}X_{\;\;\;\;\;\nu_{1}...\nu_{l}}^{\lambda\mu_{2}...\mu_{k}}+R_{\;\lambda\rho\sigma}^{\mu_{2}}X_{\;\;\;\;\;\;\nu_{1}...\nu_{l}}^{\mu_{1}\lambda\mu_{3}...\mu_{k}}+...
		\\&
		-R_{\;\nu_{1}\rho\sigma}^{\lambda}X_{\;\;\;\;\;\lambda...\nu_{l}}^{\mu_{1}...\mu_{k}}-R_{\;\nu_{2}\rho\sigma}^{\lambda}X_{\;\;\;\;\;\nu_{1}\lambda\nu_{3}...\nu_{l}}^{\mu_{1}...\mu_{k}}-...
\el\,,
 \]
we find that all terms cancel and thus $\nabla^{\beta}T^{2}_{\alpha\beta}=0$ as required.

\subsection{$S_3$}
Similarly for 
\[
S_{3}=\int d^{4}x\sqrt{-g}\left(C^{\mu\nu\lambda\sigma}\bar{{\cal F}}_{2}(\Box)C_{\mu\nu\lambda\sigma}\right)
\]
 we have the equation of motion
 \[
 \begin{aligned}T_{\alpha\beta}^{3} & =-\frac{1}{2}g_{\alpha\beta}C^{\mu\nu\lambda\sigma}\bar{{\cal F}_{3}}(\Box)C_{\mu\nu\lambda\sigma}+2C_{\alpha\mu\nu\sigma}\bar{{\cal {\cal F}}_{3}}(\square)C_{\beta}^{\;\mu\nu\sigma}-2\left(R_{\mu\nu}+2\nabla_{\mu}\nabla_{\nu}\right)\bar{{\cal {\cal F}}_{3}}(\square)C_{\beta\alpha}^{\;\;\mu\nu}\\
 & +\Theta_{\alpha\beta}^{3}-\frac{1}{2}g_{\alpha\beta}\left(\Theta_{\sigma}^{3\sigma}+\bar{\Theta}^{3}\right)+4{\cal E}_{\alpha\beta}^{3}\,.
\end{aligned}
 \]
Take the covariant derivative
\[
\begin{aligned}\nabla^{\beta}T_{\alpha\beta}^{3} & =-\frac{1}{2}\nabla_{\alpha}C^{\mu\nu\lambda\sigma}\bar{{\cal F}_{3}}(\Box)C_{\mu\nu\lambda\sigma}-\frac{1}{2}C^{\mu\nu\lambda\sigma}\nabla_{\alpha}\bar{{\cal F}_{3}}(\Box)C_{\mu\nu\lambda\sigma}+2\nabla^{\tau}C_{\alpha\mu\nu\sigma}\bar{{\cal {\cal F}}_{3}}(\square)C_{\tau}^{\;\mu\nu\sigma}\\
 & +2C_{\alpha\mu\nu\sigma}\nabla^{\tau}\bar{{\cal {\cal F}}_{3}}(\square)C_{\tau}^{\;\mu\nu\sigma}-2\left(\nabla^{\tau}R_{\mu\nu}+2\nabla^{\tau}\nabla_{\mu}\nabla_{\nu}\right)\bar{{\cal {\cal F}}_{3}}(\square)C_{\tau\alpha}^{\;\;\mu\nu}\\
 & +\nabla^{\tau}\Theta_{\alpha\tau}^{3}-\frac{1}{2}\nabla_{\alpha}\Theta_{\tau}^{3\tau}-\frac{1}{2}\nabla_{\alpha}\bar{\Theta}^{3}+4\nabla^{\tau}{\cal E}_{\alpha\tau}^{3}\,,
\end{aligned}
 \]
Again, using \eqref{tensors}, we solve for
\[
\nn
\nabla^{\tau}\Theta_{\alpha\tau}^{3}=\sum_{n=1}^{\infty}f_{3_{-n}}\sum_{l=0}^{n-1}\left[\nabla^{\tau}\nabla_{\alpha}C_{\;\nu\lambda\sigma}^{\mu(-l-1)}\nabla_{\tau}C_{\mu}^{\;\nu\lambda\sigma(l-n)}+\nabla_{\alpha}C_{\;\nu\lambda\sigma}^{\mu(-l-1)}C_{\mu}^{\;\nu\lambda\sigma(-n+l+1)}\right]\,,
   \]
\[
\nn
-\frac{1}{2}\nabla_{\alpha}\Theta_{\tau}^{3\tau}=\sum_{n=1}^{\infty}f_{3_{-n}}\sum_{l=0}^{n-1}\left[-\nabla_{\alpha}\nabla_{\tau}C_{\;\nu\lambda\sigma}^{\mu(-l-1)}\nabla^{\tau}C_{\mu}^{\;\nu\lambda\sigma(l-n)}\right]\,,
   \]
\[
\nn
-\frac{1}{2}\nabla_{\alpha}\bar{\Theta}^{3}=\sum_{n=1}^{\infty}f_{3_{-n}}\sum_{l=0}^{n-1}\left[-\nabla_{\alpha}C_{\;\nu\lambda\sigma}^{\mu(-l-1)}C_{\mu}^{\;\nu\lambda\sigma(-n+l+1)}\right]\,,
   \]
\[
\nabla^{\tau}{\cal E}_{\alpha\tau}^{3}=\sum_{n=1}^{\infty}f_{3_{-n}}\sum_{l=0}^{n-1}\biggl[[\nabla^{\tau},\nabla_{\nu}]C_{\;\;\;\sigma\mu}^{\lambda\nu(-l-1)}\nabla_{\alpha}C_{\lambda\tau}^{\;\;\;\sigma\mu(l-n)}+[\nabla_{\nu},\nabla^{\tau}]\nabla_{\alpha}C_{\;\;\sigma\mu}^{\lambda\nu(l-n)}C_{\lambda\tau}^{\;\;\sigma\mu(-l-1)}\,.
   \]
Finally using the general formula given in \eqref{covcom}, we find that all terms cancel and thus $\nabla^{\beta}T^{3}_{\alpha\beta}=0$ as required.

\section{Weak-Field Limit of Weyl Tensor}
\label{sec:weyl}
\[
\bl
C^{\beta\mu\nu\alpha}   &=      \frac{1}{2}(\partial^{\nu}\partial^{\mu}h^{\beta\alpha}+\partial^{\alpha}\partial^{\beta}h^{\mu\nu}-\partial^{\nu}\partial^{\beta}h^{\mu\alpha}-\partial^{\alpha}\partial^{\mu}h^{\beta\nu})
\\&     -       \frac{1}{4}g^{\beta\nu}\left(\partial^{\sigma}\partial^{\mu}h_{\sigma}^{\alpha}+\partial^{\alpha}\partial^{\sigma}h_{\sigma}^{\mu}-\Box h^{\mu\alpha}-\partial^{\alpha}\partial^{\mu}h\right)+\frac{1}{4}g^{\beta\alpha}\left(\partial^{\sigma}\partial^{\mu}h_{\sigma}^{\nu}+\partial^{\nu}\partial^{\sigma}h_{\sigma}^{\mu}-\Box h^{\mu\nu}-\partial^{\nu}\partial^{\mu}h\right)
        \\&-    \frac{1}{4}\left(\partial^{\sigma}\partial^{\beta}h_{\sigma}^{\nu}+\partial^{\nu}\partial^{\sigma}h_{\sigma}^{\beta}-\Box h^{\beta\nu}-\partial^{\nu}\partial^{\beta}h\right)g^{\mu\alpha}+\frac{1}{4}\left(\partial^{\sigma}\partial^{\beta}h_{\sigma}^{\alpha}+\partial^{\alpha}\partial^{\sigma}h_{\sigma}^{\beta}-\Box h^{\alpha\beta}-\partial^{\alpha}\partial^{\beta}h\right)g^{\mu\nu}
        \\&+    \frac{1}{6}\left(\partial^{\tau}\partial^{\sigma}h_{\sigma\tau}-\square h\right)g^{\beta\nu}g^{\mu\alpha}-\frac{1}{6}\left(\partial^{\tau}\partial^{\sigma}h_{\sigma\tau}-\square h\right)g^{\beta\alpha}g^{\mu\nu}
 \el
 \]
\section{$R\Box^{-2}R$-Model Contour Integrals}
\label{sec:appCI}
For the action \eqref{massive2} we read off
\[
{\cal F}_{1}(\Box)\Box=\frac{1}{3}\frac{M^{2}}{\Box},\qquad{\cal F}_{2}(\Box)={\cal F}_{3}(\Box)=0
\] 

so that
\[
\bar{a}(\Box)=1,\qquad\bar{c}(\Box)=\frac{3\Box-2M^{2}}{3\Box}
 \]

i.e.
\[
\bar{a}(\Box)=1,\qquad\bar{c}(\Box)=\frac{3p^{2}+2M^{2}}{3p^{2}}
 \]
We note the contrary to \ref{sec:MG}, in this case $a\neq c$. The Newtonian potentials then become
\[
\bl
\Phi(r)	&=	-\frac{m}{12\pi^{2}M_{P}^{2}r}\int_{0}^{\infty}dp\frac{(3p^{2}+4M^{2})\sin(pr)}{p(p^{2}+M^{2})}
	\\&=	-\frac{m}{12\pi^{2}M_{P}^{2}r}\frac{1}{4i}\int_{-\infty}^{\infty}dp\frac{(3p^{2}+4M^{2})(e^{ipr}-e^{-ipr})}{p(p+iM)(p-iM)}
 \el
 \nn
 \]
\[
\bl
\Psi(r)	&=	-\frac{m}{12\pi^{2}M_{P}^{2}r}\int_{0}^{\infty}dp\frac{(3p^{2}+2M^{2})\sin(pr)}{p(p^{2}+M^{2})}
	\\&=	-\frac{m}{12\pi^{2}M_{P}^{2}r}\frac{1}{4i}\int_{-\infty}^{\infty}dp\frac{(3p^{2}+2M^{2})(e^{ipr}-e^{-ipr})}{p(p+iM)(p-iM)}
 \el
 \]

We use the general contour integral formula given in \eqref{contform} and begin by considering $\Phi(r)$ and the pole at $p=iM$  on the upper plane. Then
\[
{\cal I}_{1}=\frac{1}{4i}\oint_{iM}dp\frac{(3p^{2}+4M^{2})e^{ipr}/p(p+iM)}{(p-iM)}=-\frac{\pi}{4}e^{-Mr}
 \]

On the lower plane $p=-iM$  we find
\[
{\cal I}_{2}=\frac{1}{4i}\oint_{-iM}dp\frac{-(3p^{2}+4M^{2})e^{-ipr}/p(p-iM)}{(p+iM)}=-\frac{\pi}{4}e^{-Mr}
 \]

Next, we consider the pole at $p=0$ by taking the limit as $\epsilon$ approaches zero  with $\epsilon>0$
 
\[
{\cal I}_{3}=\lim_{\epsilon\rightarrow0}\frac{1}{4i}\oint_{i\epsilon}dp\frac{(3p^{2}+4M^{2})e^{ipr}/(p+iM)(p-iM)}{(p-i\epsilon)}=\lim_{\epsilon\rightarrow0}\frac{2\pi i}{4i}\frac{(-3\epsilon^{2}+4M^{2})e^{-\epsilon r}}{(i\epsilon+iM)(i\epsilon-iM)}=2\pi
 \]

Combining these, we find the Newtonian potential to be
\[
\Phi(r)=-\frac{m}{12\pi^{2}M_{P}^{2}r}\left(-\frac{\pi}{2}e^{-Mr}+2\pi\right)=\frac{m\left(e^{-Mr}-4\right)}{24\pi M_{P}^{2}r}
 \]

Similarly for the $\Psi$-Integral: at the pole $p=iM$
 \[
{\cal I}_{4}=\frac{1}{4i}\oint_{iM}dp\frac{(3p^{2}+2M^{2})e^{ipr}/p(p+iM)}{(p-iM)}=\frac{\pi}{4}e^{-Mr}
 \]

at the pole $p=-iM$
 \[
{\cal I}_{5}=\frac{1}{4i}\oint_{iM}dp\frac{-(3p^{2}+2M^{2})e^{-ipr}/p(p-iM)}{(p+iM)}=\frac{\pi}{4}e^{-Mr}
 \]

at the pole $p=0$
 \[
{\cal I}_{3}=\lim_{\epsilon\rightarrow0}\frac{1}{4i}\oint_{i\epsilon}dp\frac{(3p^{2}+2M^{2})e^{ipr}/(p+iM)(p-iM)}{(p-i\epsilon)}=\lim_{\epsilon\rightarrow0}\frac{2\pi i}{4i}\frac{(-3\epsilon^{2}+2M^{2})e^{-\epsilon r}}{(i\epsilon+iM)(i\epsilon-iM)}=\pi
 \]

So that 
\[
\Psi(r)=-\frac{m}{12\pi^{2}M_{P}^{2}r}\left(\frac{\pi}{2}e^{-Mr}+\pi\right)==-\frac{m\left(e^{-Mr}+2\right)}{24\pi M_{P}^{2}r}
 \]

\bibliography{IRrefs}

\end{document}